\pdfoutput=1

\documentclass[11pt,twoside,a4paper,cmspaper,final,collab]{cms-tdr}
\def\svnVersion{464697P}\def\cmsCernNoTag{CERN-EP-2017-338}\def\cmsCernDate{\today}\def\cmsMessage{Published in the Journal of High Energy Physics as \href{http://dx.doi.org/10.1007/JHEP06(2018)031}{\doi{10.1007/JHEP06(2018)031.}}}

\begin{document}\cmsNoteHeader{B2G-17-009}

\hyphenation{had-ron-i-za-tion}
\hyphenation{cal-or-i-me-ter}
\hyphenation{de-vices}
\RCS$HeadURL: svn+ssh://svn.cern.ch/reps/tdr2/papers/B2G-17-009/trunk/B2G-17-009.tex $
\RCS$Id: B2G-17-009.tex 464645 2018-06-15 07:24:02Z grauco $

\newlength\cmsFigWidth
\setlength\cmsFigWidth{0.4\textwidth}
\providecommand{\cmsLeft}{left\xspace}
\providecommand{\cmsRight}{right\xspace}

\newcommand{\intLumi}{\ensuremath{35.9\fbinv}\xspace}
\newcommand{\hT}{\HT}
\newcommand{\mB}{\ensuremath{m_{\PB}}}
\newcommand{\MB}{\ensuremath{m_{\cPqb\PH}}}
\newcommand{\mJ}{\ensuremath{m_{\text J}}}
\newcommand{\BHb}{\ensuremath{\PB\to\PH\PQb}\xspace}

\providecommand{\CL}{CL\xspace}
\newlength\cmsTabSkip\setlength{\cmsTabSkip}{1ex}

\cmsNoteHeader{B2G-17-009}
\title{Search for single production of vector-like quarks decaying to a \PQb quark and a Higgs boson}

\date{\today}

\abstract{A search is presented for single production of heavy vector-like quarks (\PB) that decay to a Higgs boson and a \PQb~quark, with the Higgs boson decaying to a highly boosted \bbbar pair reconstructed as a single collimated jet.
The analysis is based on data collected by the CMS experiment in proton-proton collisions at  $\sqrt{s} = 13\TeV$, corresponding to an integrated luminosity of \intLumi. The data are consistent with background expectations, and upper limits at 95\% confidence level on the product of the \PB~quark cross section and the branching fraction are obtained in the range 1.28--0.07\unit{pb}, for a narrow \PB~quark with a mass between 700 and 1800\GeV. The production of \PB~quarks with widths of 10, 20 and 30\% of the resonance mass is also considered, and the sensitivities obtained are similar to those achieved in the narrow width case.
This is the first search at the CERN LHC for the single production of a \PB~quark through its fully hadronic decay channel, and the first study considering finite resonance widths of the \PB~quark.}

\hypersetup{
pdfauthor={CMS Collaboration},
pdftitle={Search for single production of vector-like quarks decaying to a b quark and a Higgs boson},
pdfsubject={CMS},
pdfkeywords={CMS, physics, vector-like quarks, beyond standard model}}

\maketitle

\section{Introduction} \label{intro}
\hyphenation{models} \hyphenation{predominant}

With the discovery of the Higgs boson (\PH) by the ATLAS~\cite{Aad:2012tfa} and CMS~\cite{Chatrchyan:2012xdj, Aad:2015zhl} experiments at the CERN LHC, the standard model (SM) of particle physics has now been completely confirmed. However, the SM does not address, for example, problems related to the nature of the electroweak symmetry breaking and the hierarchy between the electroweak and the Planck mass scales. Several extensions of the SM address such issues through the introduction of new particles that allow the cancellation of loop corrections to the mass of the Higgs boson \cite{PhysRevD.88.094010}.
Supersymmetric theories propose bosonic partners of the top quark to address the hierarchy problem; other models such as Little Higgs or Composite Higgs boson models~\cite{Schmaltz:2005ky, Kaplan:1983fs, Georgi:1984ef, Dugan:1984hq} overcome the hierarchy problem by introducing heavy fermionic resonances called vector-like quarks (VLQs)~\cite{PhysRevD.88.094010, 1126-6708-2009-11-030,  DeSimone2013, Buchkremer2013376}.
The vector-like nature of these quarks does not exclude their having a fundamental mass, in contrast to chiral fermions, which acquire mass via electroweak symmetry breaking in the SM. The VLQs are therefore not excluded by present searches, unlike a fourth generation of SM quarks that is ruled out by electroweak precision measurements~\cite{Kribs:2007nz, Banerjee:2013hxa}, and by the measured properties of the SM Higgs boson ~\cite{Khachatryan:2016vau, PhysRevD.86.013011, 201336}. Previous searches for VLQs have been performed by the ATLAS~\cite{PhysRevD.91.112011, PhysRevD.92.112007, Aad:2014efa, Aad:2016shx, Aad2016, Aad20162} and CMS~\cite{PhysRevD.93.012003, Khachatryan:2015gza, PhysRevLett.112.171801, Sirunyan:2017ezy, Khachatryan:2016vph, CMS-PAS-B2G-16-005, CMS-PAS-B2G-16-006} experiments in proton-proton collisions recorded at centre-of-mass energies of 7, 8, and 13\TeV.

\begin{figure*}[hbtp]
	\centering
    	\includegraphics[width=0.65\textwidth]{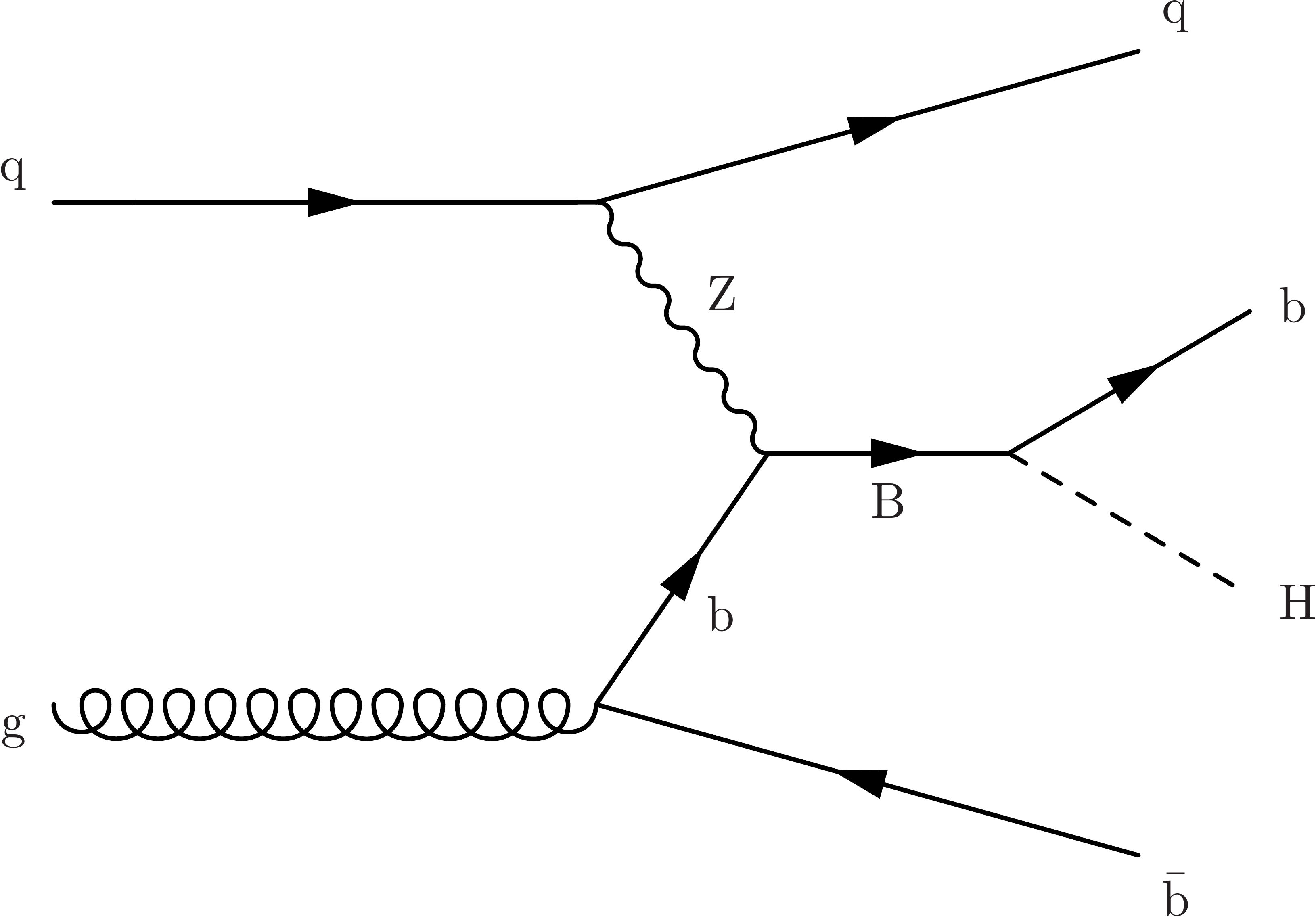}
    	\caption{The leading-order Feynman diagram for the production of a single vector-like \PB~quark in association with a \PQb~quark and light-flavour quark, and its decay to a Higgs boson and a \PQb~quark.}
	\label{fig:BprimeBToBH}
\end{figure*}

We present a search for electroweak production of single vector-like \PB~quarks with electrical charge $-1/3\,e$, with $e$ the proton charge, that decay to a bottom (\PQb) quark and a Higgs boson. The search uses \Pp\Pp~events collected by the CMS experiment at a centre-of-mass energy of 13\TeV, corresponding to an integrated luminosity of \intLumi. We study the fully hadronic final state with the Higgs boson decaying to a pair of \PQb~quarks. Figure~\ref{fig:BprimeBToBH} illustrates the electroweak production of a \PB~quark in association with a \PQb and a light-flavour quark, typically emitted into the forward region of the detector.

The \PB~decay channel considered in this analysis is \BHb. However, the \PB~quark can also decay into \cPZ\PQb, \PW\PQt, and possibly into lighter states predicted in models beyond the SM that have model-dependent branching fractions. Our results are interpreted assuming that the \PB~quark belongs to a singlet or doublet representation and that it decays exclusively to SM particles.
The singlet branching fractions of the \PB~quark into \PH{}\PQb, \cPZ\PQb, and \PW\PQt are $\mathcal{B} \approx 25$, 25, and 50\%, and the doublet branching fractions are 50, 50, and 0\%, and all depend on the vector-like quark mass \mB.

Previous CMS searches for vector-like \PB~quarks relied on the assumption of a decay width that is narrow compared to the experimental resolution. The present analysis, in addition to searching for \PB~quarks with narrow decay widths, also explores the possibility that \PB~quarks have a non-negligible width, with values up to 30\% of the resonance mass. In comparison, the experimental resolution in the reconstructed \PB~mass, defined as the ratio between the root-mean-square width of the peak and its mean position, ranges between 8 and 15\%, depending on the mass hypothesis. In addition to broadening the width of the observed signal, the intrinsic width of the resonance would modify the kinematic distributions of the final state, thus changing the selection efficiency. These effects are taken into account in this analysis.

The cross section for single production of a \PB~quark depends on \mB~and its electroweak couplings to SM particles. The kinematic distributions depend only on the total width of the \PB~quark. The benchmark model in this analysis assumes a weak coupling of the \PB~quark to the \cPZ~boson and \PQb~quark. Because of the mixing between \PB~and the SM bottom quark in models where \PB~is a singlet or part of a doublet, the \PB\cPqb\cPZ~electroweak coupling has a predominant chirality, respectively, right- or left-handed. The coupling chirality can potentially affect the kinematic distributions. We explicitly checked and found that these effects are negligible for the channel discussed in this work, and our results can therefore be interpreted in both singlet and doublet models.

\section{The CMS detector and particle reconstruction}\label{detector}
\hyphenation{response}

The central feature of the CMS apparatus is a superconducting solenoid of 6\unit{m} internal diameter, providing a magnetic field of 3.8\unit{T}. A silicon pixel and strip tracker, a lead tungstate crystal electromagnetic calorimeter (ECAL), and a brass and scintillator hadron calorimeter (HCAL), each composed of a barrel and two end sections, reside within the solenoid. Forward calorimeters extend the pseudorapidity ($\eta$) coverage provided by the barrel and end detectors. Muons are measured in gas-ionization detectors embedded in the steel flux-return yoke outside the solenoid. A more detailed description of the CMS detector, together with a definition of the coordinate system and kinematic variables, can be found in Ref.~\cite{Chatrchyan:2008aa}.

Events of interest are selected using a two-tiered trigger system~\cite{Khachatryan:2016bia}. The first level, composed of specialized hardware processors, uses information from the calorimeters and muon detectors to select events at a rate of $\approx$100\unit{kHz} within a time interval of less than 4\unit{\mus}. The second level, known as the high-level trigger (HLT), consists of a farm of processors running a version of the full event-reconstruction software optimized for fast processing that reduces the event rate to $\approx$1\unit{kHz} before data storage.

Event reconstruction is based on the CMS particle-flow (PF) algorithm \cite{CMS-PRF-14-001}, which reconstructs and identifies each individual particle through an optimized combination of information from the various elements of the CMS detector. The energy of electrons is defined through the combination of the electron momentum at the primary interaction vertex determined in the tracker, the energy of the corresponding ECAL cluster, and the energy sum of all bremsstrahlung photons spatially compatible with originating from the electron track from the primary \Pp\Pp~collision vertex. The energy of muons is obtained from the curvature of the corresponding track. The reconstructed energy of charged hadrons is extracted from the reconstructed tracks in the tracker and their matching energy depositions in ECAL and HCAL. Energy depositions are corrected for ignoring calorimeter readouts that are close to threshold (zero suppression) and for the response function of calorimeters to hadronic showers. Finally, the energy of neutral hadrons is obtained from the corresponding corrected ECAL and HCAL energies.

Jets are reconstructed by clustering PF candidates using the anti-\kt algorithm \cite{Cacciari:2008gp}. Prior to clustering, the charged-hadron subtraction algorithm \cite{CMS:2014ata} is applied to the event to reduce the effects of pileup (i.e. additional \Pp\Pp~collisions occurring within the same or neighbouring LHC bunch crossings).

This algorithm discards charged hadrons not originating from the primary vertex, which is defined as the reconstructed vertex with the largest value of summed $\pt^2$ of charged hadrons contributing to jets. The jets are clustered using the jet finding algorithm~\cite{Cacciari:2008gp,Cacciari:2011ma}, which defines the associated missing transverse momentum taken as the negative of the vector sum of the \pt of those jets.  We consider jets with different distance parameter of $\Delta R = \sqrt{\smash[b]{(\Delta y)^2+(\Delta\phi)^2}} = 0.4$ or 0.8, with $y$ the rapidity, referred to as AK4 and AK8 jets, respectively.

The residual pileup contamination from neutral hadrons is subtracted, assuming that it is proportional to the event energy density and the jet area, estimated using the {\FASTJET} package \cite{Cacciari:2011ma}. Jet momenta are determined from the vectorial sum of all the individual PF particles in the jet. The energy scale calibrations obtained from Monte~Carlo (MC) simulation are applied to correct the four-momenta of jets. Residual corrections, accounting for remaining discrepancies between jet response in data and in simulated events, are applied to the former. The jet energy resolution for simulated jets is degraded slightly to reproduce the resolution in data. The AK4 jet candidates are required to have $\pt > 30\GeV$ and  $\abs{\eta}<4$, and to satisfy a stringent set of identification criteria designed to reject spurious detector and reconstruction effects \cite{CMS-PAS-JME-16-003}. The jets with  $\abs{\eta}>2.4$ are referred to as forward jets. The AK8 jets, used to identify and reconstruct Higgs boson candidates, are selected to have $\pt > 300\GeV$ and  $\abs{\eta}<2.4$.

A multivariate \PQb tagging algorithm (CSVv2) \cite{BTV-16-002} is used to identify central jets (with $\abs{\eta}<2.4$) arising from the hadronization of \PQb~quarks. Parameters are chosen for the CSVv2 discriminant such that the tagging efficiency for b quark jets is $\approx$70\% while the identification probability averaged over the jet kinematics in \ttbar events is $\approx$1\% for light flavour jets with $\pt > 30\GeV$.

The Higgs boson candidates are identified using the heavy-flavour content of the AK8 jet. A pruning algorithm \cite{PhysRevD.80.051501} is applied that uses the Cambridge--Aachen (CA) algorithm \cite{CATANI1993187} to recluster each AK8 jet starting from all its original constituents and to discard soft and wide-angle radiation inside the jet in each step of the iterative procedure. The procedure defines a pruned-jet mass, computed from the sum of the four-momenta of the constituents that have not been removed by the pruning algorithm, which achieves a better mass resolution. The pruned mass of the jet is then used as a discriminant to reject quark and gluon jets and to select Higgs bosons, by requiring its mass to be within the window of 105--135\GeV. Two subjets are obtained using the soft drop declustering algorithm \cite{Dasgupta:2013ihk, Larkoski:2014wba}, and these are required to pass the same CSVv2 discriminant threshold used for the AK4 jets.

\section{Modelling and simulation}\label{samples}

The production and decay of high mass \BHb, with \PH~\ensuremath{\to}~\bbbar, provides a signature with multiple jets rich in heavy-flavour content, and characterized by a highly boosted Higgs boson. The dominant background in this search is from SM events comprised of jets produced through the strong quantum chromodynamic (QCD) interaction, referred to as multijet events. Additional contributions arise from \ttbar events, and minor backgrounds are associated with the production of \PW~or \cPZ~bosons in association with jets.

Simulated events are used throughout the analysis to define selection strategy and to determine the expected sensitivity to vector-like quarks. The background from multijet events is estimated using data in control regions. Simulation is also used to cross-check the multijet background prediction and to evaluate its validity. The contributions from other backgrounds, such as \ttbar events and \PW~or \cPZ~boson production in association with jets, are estimated through MC simulation.

Multijet events, as well as electroweak backgrounds from virtual or on-mass shell \cPZ~or $\gamma$+jets and \PW+jets production, are simulated at leading order (LO) using the \MGvATNLO 2.2.2 generator \cite{Alwall:2014hca}, interfaced to \PYTHIA~8.2~\cite{pythia8} with the CUETP8M1~\cite{Skands2014, cuet} underlying-event tune for parton-shower simulation and evolution. The background \ttbar events are generated using \POWHEG~v2 at next-to-leading order (NLO)~\cite{Nason:2004rx, Frixione:2007nw, Frixione:2007vw, Alioli:2010xd}, also interfaced to \PYTHIA. The mass of the top quark is set to 172.5\GeV, and the cross section is calculated at next-to-next-to-leading order (NNLO) in perturbative QCD using a next-to-next-to-leading-logarithmic (NNLL) soft-gluon approximation (NNLO+NNLL) in the \textsc{Top++} 2.0 program~\cite{CZAKON20142930}. The cross sections for \cPZ~or $\gamma$+jets and \PW+jets processes are calculated at NNLO using the \FEWZ~MC program \citep{Li:2012wna}.

The \BHb~\ensuremath{\to}~{\bbbar}\PQb~events are simulated at LO, modelled using the universal \textsc{FeynRules} output \cite{Degrande:2011ua, Matsedonskyi:2014mna} and the MC generator \MGvATNLO, interfaced to \PYTHIA~8 for parton-shower simulation. Several mass hypotheses are considered for signals in the range $700 < \mB < 1800\GeV$, in steps of 100\GeV for total decay widths of 1\GeV, representing the narrow-width categories. Signal events for \PB~quarks with large widths (10, 20, or 30\% of the mass hypothesis) are also generated in the same mass range. All \PB~quarks are generated with left-handed chirality, but the effect on the kinematic distributions of only considering one chirality is found to be negligible. Interference between the signal and the SM background is negligible.

Simulations using LO and NLO calculations, respectively, use the LO and NLO~{NNPDF3.0}~\cite{Ball:2017nwa} sets of parton distribution functions (PDFs). All signal and background events are processed using \GEANT4 \cite{AGOSTINELLI2003250} to provide a full simulation of the CMS detector. The generated events are also reweighted to account for the dependence of the reconstruction efficiency on the number of pileup interactions in the collisions.

\section{Interpretation framework}\label{theory}

The total cross section for the single production and decay of a \PB~quark with final state X can be written as:
\begin{equation}
	\label{eq:AW}
	\sigma(C_1,C_2,\mB,\Gamma_{\PB},X) = C_1^2~C_2^2~\hat\sigma_\text{AW}(\mB,\Gamma_{\PB}),
\end{equation}
where $C_1$ and $C_2$ are the production and decay couplings corresponding to the interactions through which a B quark is produced and decays, and $\hat\sigma_\text{AW}$ is the reduced cross section for a resonance of arbitrary width (AW). This width can be written as $\Gamma_{\PB} = \Gamma (C_i,\mB,m_{\text{decays}})$, as it depends on the B quark mass, on the masses of all its decay products, and on its couplings to all decay channels, $C_i$.

Equation~(\ref{eq:AW}) is valid in all width regimes. However, when $\Gamma_{\PB}/$\mB~ approaches zero, it is possible to factorize production and decay and to write the cross section as:
\begin{equation}
	\label{eq:NWA}
	\sigma(C_1,C_2,\mB,\Gamma_{\PB}) = \sigma_\text{prod}(C_1,\mB)\mathcal{B}_{{\PB} \to {\text{X}}} = C_1^2\hat\sigma_{\text{NWA}}(\mB)\mathcal{B}_{{\PB} \to {\text{X}}},
\end{equation}
where $C_1$ is the \PB~production coupling, and information for the parameters $C_2$ and $\Gamma_{\PB}$ are included in the branching fraction for the specific \PB~quark decay, in this case $\mathcal{B}_{{\PB} \to {\text{X}}}$, while $\hat\sigma_{\text{NWA}}(\mB)$ is the reduced cross section in the narrow-width approximation (NWA).

Our assumptions have the \PB~quark decaying into \PH{}\PQb, \cPZ\PQb, and \PW\PQt~with branching fractions that are specified in the model.
The couplings of the \PB~quark to SM bosons and quarks can be parametrized as:
$\text{c}_{\cPZ} = {e}/({2 \text{c}_\text{w} \text{s}_\text{w}}\kappa_{\cPZ})$, $\text{c}_{\PW} = e /({\sqrt{2} \text{s}_\text{w}}\kappa_{\PW})$, and $\text{c}_{\PH} = (m_{\PB}\kappa_{\PH})/{v}$, where $e$ is the electric charge of the proton, $v = 246$\GeV is the vacuum-expectation value for the field of the Higgs boson, $\text{c}_\text{w}$ and $\text{s}_\text{w}$ are the cosine and sine of the weak mixing angle $\theta_{\PW}$, and $\kappa$ is a coupling strength that can be fixed to obtain the desired width.
Numerically, $ {e}/({2 \text{c}_\text{w} \text{s}_\text{w}}) = 0.370$, and $\text{c}_{\PW} = {e}/({\sqrt{2} \text{s}_\text{w}}) = 0.458$.
For the process under consideration, we can set $C_1\equiv c_{\cPZ}$ and $C_2\equiv c_{\PH}$.

The $\kappa$ values can be related to the mixing angle between the vector-like \PB~quark and the \PQb~quark~\cite{Chen:2017hak}, and correspond to left- and right-handed couplings, which are the dominant chiralities for a singlet or part of a doublet \PB~quark, respectively.
For small values of $\kappa$, corresponding to the NWA regime, the following relations hold to excellent approximation: for a \PB~singlet $\kappa_{\cPZ}\approx\kappa_{\PH}\approx\kappa_{\PW}\approx\kappa$, while for a (T,\PB) doublet (where T~is a vector-like quark with electrical charge 2/3) with no vector-like top quark Yukawa coupling, $\kappa_{\cPZ}\approx\kappa_{\PH}\approx\kappa$, and $\kappa_{\PW} = 0$.
By imposing these relations among the $\kappa$ values, and fixing the $\Gamma_{\PB}/$\mB~ratio to 1\%, $\kappa$ is $\approx$0.1 in the whole range of explored masses.
Table~\ref{tab:NWAsigmafixkappa} provides the values for $\hat{\sigma}_{\text{NWA}}$ and the physical cross sections in the NWA for the $\Pp\Pp\to\PB\cPqb\cPq$ process. The CTEQ6L PDF set~\cite{Pumplin:2002vw} is used in this calculation.

To interpret the results in a model-independent way, the mechanism through which the B quarks achieve large widths is not specified, and $\Gamma_{\PB}$ is considered as a free parameter. The relations among the $\kappa_\text{X}$ (with X = {\PW}, {\cPZ}, {\PH}), corresponding to the NWA limit ($\kappa_{\cPZ} = \kappa_{\PH} = \kappa_{\PW} = \kappa$), are imposed for the large-width regime.
 With this assumption, the total width $\Gamma_{\PB}$ is always proportional to $\kappa^2$, and therefore $\kappa$ can be chosen to obtain a specific $\Gamma_{{\PB}}/$\mB~ratio. However, with the assumption relaxed, in a simplified model, new physics can be invoked to generate the required couplings.

\begin{table*}[hbtp]
	\centering
	{
	\setlength{\tabcolsep}{2.5pt}
	\topcaption{ \label{tab:NWAsigmafixkappa} Cross sections for $\Pp\Pp\to\PB\cPqb\cPq$, with the ratio $\Gamma_{\PB}/$\mB~fixed to 1\% (NWA). The couplings and branching fractions in simplified models are calculated using the equations in the text. The uncertainties in the production cross sections correspond to the halving and doubling of the QCD renormalization and factorization scales}

	\begin{tabular}{cc|ccccc|cccc}
		\hline
		 & & \multicolumn{5}{c|}{Singlet model} &  \multicolumn{4}{c}{Doublet model} \\
		\mB~(\GeVns) & $\hat{\sigma}_\text{NWA}$ (\unit{pb}) & $\kappa$ & $\mathcal{\PB}_{{\PB} \to {\PW\PQt}}$ & $\mathcal{B}_{{\PB} \to {\cPZ\PQb}}$  & $\mathcal{B}_{{\PB} \to {\PH\PQb}}$ & $\sigma_\text{NWA}$ (\unit{pb}) & $\kappa$ & $\mathcal{B}_{{\PB} \to {\cPZ\PQb}}$  & $\mathcal{B}_{{\PB} \to {\PH\PQb}}$ & $\sigma_\text{NWA}$ (\unit{pb})  \\
		\hline\\[-0.8em]
		700  & 31.30  $^{+28\%}_{-20\%}$ & 0.18 & 0.466 & 0.271 & 0.263 & 0.1631 & 0.25 & 0.499 & 0.501 & 0.5720 \\
		800  & 21.50  $^{+29\%}_{-21\%}$ & 0.16 & 0.474 & 0.276 & 0.260 & 0.0830 & 0.22 & 0.499 & 0.501 & 0.3003 \\
		900  & 15.10  $^{+30\%}_{-21\%}$ & 0.14 & 0.489 & 0.263 & 0.258 & 0.0451 & 0.19 & 0.500 & 0.500 & 0.1666 \\
		1000 & 10.80  $^{+31\%}_{-23\%}$ & 0.13 & 0.483 & 0.261 & 0.256 & 0.0257 & 0.17 & 0.500 & 0.500 & 0.0962 \\
		1100 & 7.85  $^{+32\%}_{-22\%}$ & 0.11 & 0.486 & 0.259 & 0.255 & 0.0153 & 0.16 & 0.500 & 0.500 & 0.0580 \\
		1200 & 5.77  $^{+33\%}_{-23\%}$ & 0.10 & 0.489 & 0.257 & 0.254 & 0.0094 & 0.15 & 0.500 & 0.500 & 0.0358 \\
		1300 & 4.29  $^{+34\%}_{-23\%}$ & 0.10 & 0.490 & 0.256 & 0.254 & 0.0059 & 0.13 & 0.500 & 0.500 & 0.0227 \\
		1400 & 3.23  $^{+34\%}_{-23\%}$ & 0.09 & 0.492 & 0.255 & 0.253 & 0.0038 & 0.12 & 0.500 & 0.500 & 0.0147 \\
		1500 & 2.45  $^{+35\%}_{-25\%}$ & 0.08 & 0.493 & 0.254 & 0.253 & 0.0025 & 0.12 & 0.500 & 0.500 & 0.0097 \\
		1600 & 1.86  $^{+36\%}_{-24\%}$ & 0.08 & 0.494 & 0.254 & 0.252 & 0.0017 & 0.11 & 0.500 & 0.500 & 0.0065 \\
		1700 & 1.44  $^{+37\%}_{-24\%}$ & 0.07 & 0.494 & 0.254 & 0.252 & 0.0011 & 0.10 & 0.500 & 0.500 & 0.0044 \\
		1800 & 1.11  $^{+37\%}_{-25\%}$ & 0.07 & 0.495 & 0.253 & 0.252 & 0.0008 & 0.10 & 0.500 & 0.500 & 0.0031 \\
		\hline
	\end{tabular}}
\end{table*}

Table~\ref{tab:AWsingletlike} reports the cross sections integrated over the phase space of \cPq~and \PQb, the particles produced  in association with the \PB~quark (see Fig.~\ref{fig:BprimeBToBH}), for fixed values of $\Gamma_{\PB}/$\mB, with configurations of $\kappa$ corresponding to singlet ($\sigma_\text{S}$) and doublet ($\sigma_\text{D}$) representations.
Given the yields for a doublet in the \cPZ\PQb~and \PH{}\PQb~decay modes, these couplings at fixed width are larger than for singlets, and as a consequence $\sigma_\text{D} > \sigma_\text{S}$.

\begin{table*}[hbtp]
	\centering
	\topcaption{\label{tab:AWsingletlike} Cross sections for $\Pp\Pp\to\PB\cPqb\cPq$ for three values of the $\Gamma_{\PB}/$\mB~ratio. The conditions assume that singlets and doublets have $\kappa_{\PW} = \kappa_{\cPZ} = \kappa_{\PH}\equiv\kappa$, $\kappa_{\PW} = 0$ and $\kappa_{\cPZ} = \kappa_{\PH}\equiv\kappa$, respectively. For each $\Gamma_{\PB}/$\mB, we provide the values of $\tilde\sigma_\text{AW}$ and of the physical cross sections for both the singlet and doublet models, $\sigma_\text{S}$ and $\sigma_\text{D}$ respectively. The uncertainties in the production cross sections correspond to the halving and doubling of the QCD renormalization and factorization scales. The values of $\kappa$ are listed in the parentheses. }

	\resizebox{\textwidth}{!}{
	\begin{tabular}{c|ccc|ccc|ccc}
		\hline
		 & \multicolumn{3}{c|}{$\Gamma_{\PB}/\mB = 10\%$} &  \multicolumn{3}{c|}{$\Gamma_{\PB}/\mB = 20\%$} &  \multicolumn{3}{c}{$\Gamma_{\PB}/\mB = 30\%$} \\[0.2em]
		\mB~(\GeVns) & $\tilde\sigma_\text{AW}$(\unit{pb}) & $\sigma_\text{S}$(\unit{fb}) ($\kappa$) & $\sigma_\text{D}$(\unit{fb}) ($\kappa$) & $\tilde\sigma_\text{AW}$(\unit{pb}) &  $\sigma_\text{S}$(\unit{fb}) ($\kappa$) & $\sigma_\text{D}$(\unit{fb}) ($\kappa$) & $\tilde\sigma_\text{AW}$(\unit{pb}) & $\sigma_\text{S}$(\unit{fb}) ($\kappa$) & $\sigma_\text{D}$(\unit{fb}) ($\kappa$) \\[0.1em]
		\hline\\[-0.8em]
		700 & 3.01 & 400 (0.588)  & 1378 (0.8010)  & 1.43 & 759 (0.832)  &  2616 (1.130)  & 0.899 & 1074 (1.020)  & 3703 (1.390) \\[1.0mm]
		800 & 2.10 & 203 (0.508)  & 726 (0.699)  & 1.00 & 386 (0.719)  &  1377 (0.9880)  & 0.634 & 552 (0.880)  & 1968 (1.210) \\[1.0mm]
		900 & 1.51 & 111 (0.448)  & 406 (0.619)  & 0.719 & 212 (0.633)  &  775 (0.876)  & 0.454 & 301 (0.776)  & 1101 (1.070) \\[1.0mm]
		1000 & 1.09 & 63.7 (0.401)  & 237 (0.556)  & 0.523 & 122 (0.567)  &  453 (0.787)  & 0.331 & 174 (0.694)  & 647 (0.964) \\[1.0mm]
		1100 & 0.807 & 38.2 (0.363)  & 144 (0.505)  & 0.386 & 73.2 (0.513)  &  276 (0.714)  & 0.246 & 105 (0.628)  & 394 (0.875) \\[1.0mm]
		1200 & 0.601 & 23.6 (0.331)  & 89.7 (0.463)  & 0.290 & 45.5 (0.468)  &  173 (0.654)  & 0.185 & 65.2 (0.574)  & 248 (0.801) \\[1.0mm]
		1300 & 0.451 & 14.9 (0.305)  & 57.1 (0.427)  & 0.220 & 29.0 (0.431)  &  111 (0.603)  & 0.141 & 41.9 (0.528)  & 160 (0.739) \\[1.0mm]
		1400 & 0.342 & 9.70 (0.283)  & 37.2 (0.396)  & 0.167 & 18.9 (0.400)  &  72.9 (0.560)  & 0.108 & 27.5 (0.489)  & 106 (0.686) \\[1.0mm]
		1500 & 0.262 & 6.42 (0.263)  & 24.9 (0.369)  & 0.129 & 12.6 (0.372)  &  48.9 (0.522)  & 0.0836 & 18.4 (0.456)  & 71.3 (0.640) \\[1.0mm]
		1600 & 0.203 & 4.34 (0.246)  & 16.9 (0.346)  & 0.101 & 8.61 (0.349)  &  33.5 (0.489)  & 0.0651 & 12.5 (0.427)  & 48.7 (0.599) \\[1.0mm]
		1700 & 0.158 & 2.99 (0.232)  & 11.6 (0.326)  & 0.0788 & 5.94 (0.328)  &  23.2 (0.460)  & 0.0514 & 8.71 (0.401)  & 34.0 (0.564) \\[1.0mm]
		1800 & 0.124 & 2.08 (0.219)  & 8.13 (0.307)  & 0.0621 & 4.16 (0.309)  &  16.3 (0.435)  & 0.0408 & 6.14 (0.379)  & 24.0 (0.532) \\[1.0mm]
		\hline
	\end{tabular}}
\end{table*}

\section{Event selection} \label{anastrategy}

This analysis searches for a Higgs boson and a bottom quark arising from the decay of a \PB~quark, and the decay of the Higgs boson into a pair of \PQb~quarks. An additional light-flavour quark, resulting from the production mechanism and produced in the forward direction (see Fig. \ref{fig:BprimeBToBH}), is also present. For values of \mB~much larger than the Higgs boson mass, the decay products of the B quark are expected to have large \pt. The two b quarks originating from the Higgs boson tend therefore to emerge very close to each other in $\eta$-$\phi$ space, resulting in a single large jet.

The data are collected through an online selection (trigger) based on jet activity \hT, defined as the scalar \pt sum of all AK4 jets with $\pt > 30\GeV$ and  $\abs{\eta}<3$. The jet activity threshold for this trigger is 900\GeV. Collisions containing at least one jet reconstructed through the HLT system with $\pt > 450\GeV$ are also selected, to increase the HLT efficiency. At the analysis level, \hT~is recalculated using AK4 jets with $\pt > 50\GeV$ and   $\abs{\eta}<2.4$, and $\hT > 950\GeV$ is required.  This offline selection corresponds to a trigger efficiency in excess of 87\%.

\begin{figure*}[hbtp]
	\centering
    	\includegraphics[width=1.6\cmsFigWidth]{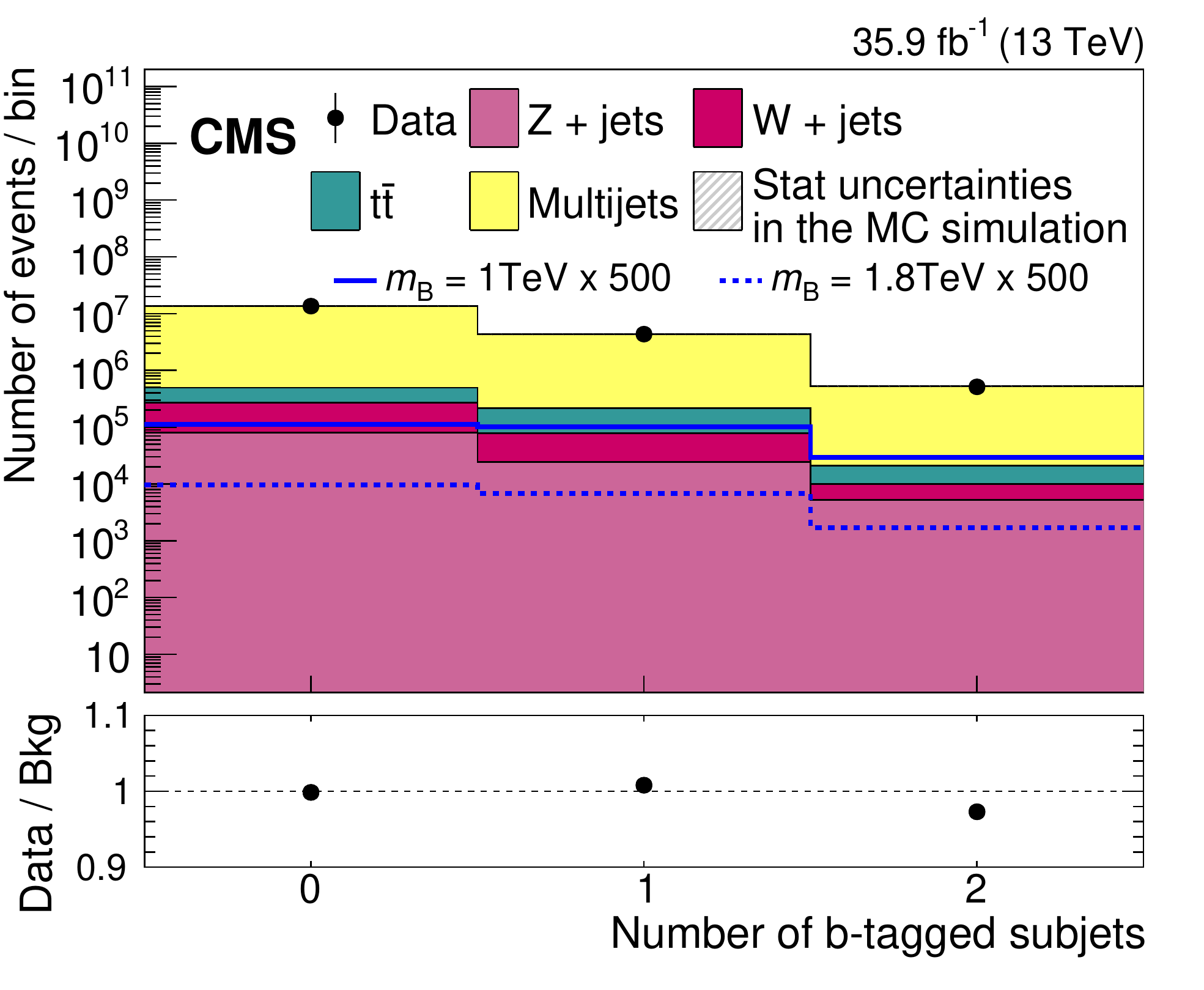}
    	\caption{The b-tagged subjet multiplicity of AK8 jets in events passing preselection criteria. The lower panel shows the ratio of data to the MC background prediction. The normalization of the contributions from signals at $\mB = 1$ and 1.8\TeV is multiplied by a factor of 500. Background events are normalized to data. Only the statistical uncertainties are taken into consideration here, and they are too small to be visible.}
    	\label{fig:HiggsTag}
\end{figure*}

Events are preselected if they contain three or more AK4 jets with $\pt > 30\GeV$ and  $\abs{\eta}<4$, among which there must be at least one b-tagged jet with  $\abs{\eta}<2.4$. A veto is applied to events with one or more leptons to ensure that the selection criteria do not overlap with those used for searches for the \PB~quark in leptonic final states. Selected events are further required to have at least one large Higgs-tagged AK8 jet, fulfilling the Higgs boson tagging requirements as described in Section~\ref{detector}. The Higgs boson tagging efficiency is 10--20\%, depending on the value of \mB. Figure~\ref{fig:HiggsTag} compares to data the b-tagged subjet multiplicity expected for simulated background and for signal processes.

The B quark is reconstructed from the Higgs jet candidate along with a nonoverlapping \PQb-tagged jet. The \PQb~quark from \PB~quark decay is usually highly energetic ($\pt > 200\GeV$), thus the \PQb~jet with the highest \pt~is chosen, and this reduces significantly the combinatorial background.  Furthermore, to reduce overlaps with the decay products of the Higgs boson, a condition is applied on the distance between the two objects in ($\eta, \phi$), requiring $\DR(\PQb,\PH) > 1.2$.

To further reduce the multijet background and the contamination from gluon-like jets, \hT~is required to be in excess of 950\GeV for smaller mass values of $700 < \mB < 1500\GeV$, while for  $1500 < \mB < 1800\GeV$, a trigger with a threshold of $\hT > 1250\GeV$ is chosen. In what follows, we refer to the former as the ``low-mass analysis'' and to the latter as the ``high-mass analysis''.

\begin{figure*}[hbtp]
	\centering
    	\includegraphics[width=1.6\cmsFigWidth]{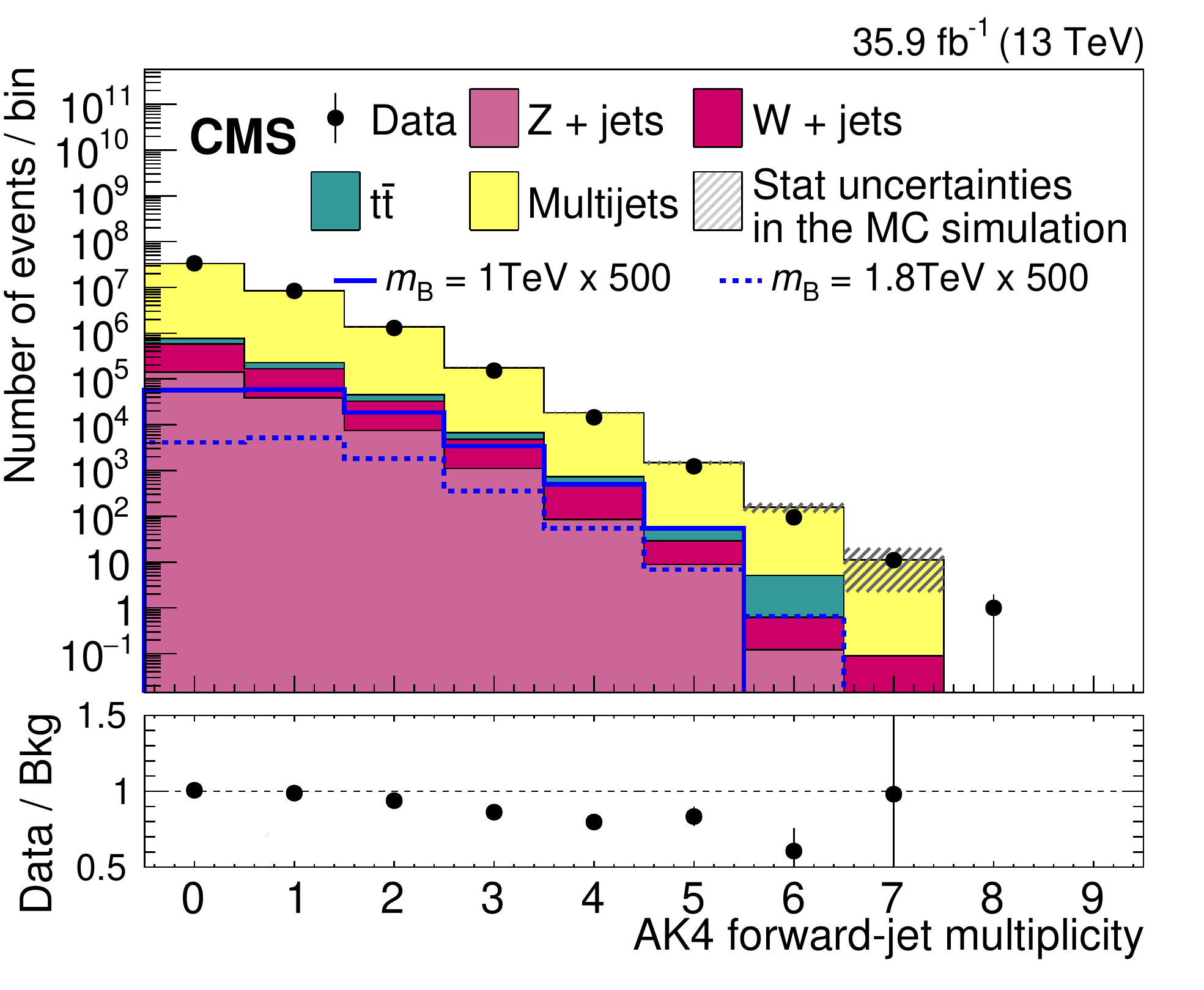}
    	\caption{Multiplicity of forward jets before event categorization. The normalization of the signal contributions is multiplied by a factor of 500. All the contributions to background are obtained from Monte Carlo simulation and are normalized to data. The lower panel shows the ratio of data to background. We show only the statistical uncertainties.}
    	\label{fig:ForwardJet}
\end{figure*}

The signal to background discrimination is enhanced by exploiting the distinctive presence of a forward jet. Events are therefore separated into categories based on the forward-jet multiplicity. A high-purity category is obtained by requiring at least one forward jet. A second category that contains a large fraction of events from both signal and background, is defined requiring no forward jets. The forward-jet multiplicity expected for background and signal events after preselection is compared to data in Fig.~\ref{fig:ForwardJet}. After all the selections are implemented, we reach signal efficiencies ranging from 2\% or less at low masses, to larger values at larger \mB, as a result of the optimization of the analysis for highly-boosted topologies.
The disagreement between data and simulation at large forward-jet multiplicities does not affect the analysis, as the background contribution in the signal region is estimated from data. Moreover, the effect on the measurement is negligible since the majority of vector-like \PB~quark events contain less than 2 forward jets, for which the simulated and observed yields are consistent after preselection.

\section{Signal extraction} \label{bkg}

A potential signal would manifest itself as a localized excess over the expected background in the spectrum of the reconstructed mass~\MB. A binned maximum likelihood fit is performed to the \MB~distribution to extract a signal, exploiting the characteristic structure of the reconstructed \PB~quark mass spectrum.

Multijet events constitute the dominant source of background in this search. An additional contribution of 5--7\% arises from \ttbar events. To reduce the dependence of the maximum-likelihood fit on the modelling of the multijet background in simulation, the contribution from this background is obtained from data. The procedure we use to estimate the yield of such events in the signal region is referred to as the ``ABCD method'' (discussed below), but its dependence on \MB~is taken from a background-enriched control region in data. A minor contribution ($\approx$1\%) to the background arises from other SM sources, such as \cPZ+jets and \PW+jets events. Both \ttbar events and these minor backgrounds are estimated from simulation. The normalization of multijet events in the signal region is estimated using three data control regions, enriched in background events. These regions, in addition to the one enriched in signal events,  are sampled in a two-dimensional phase space defined by two variables: the b-tagged subjet multiplicity of the Higgs jet and its reconstructed mass, \mJ. From a check on the simulation, the number of b-tagged subjets is not correlated with \mJ. The four regions used to define the ABCD method are: (i) region A, with two  b-tagged subjets, and $105 < \mJ < 135$\GeV, (ii) region B, with two b-tagged subjets, and $75 < \mJ < 105\GeV$ or $\mJ > 135\GeV$, (iii) region C, with one b-tagged subjet, and $105 < \mJ < 135\GeV$, and (iv) region D, with one  b-tagged subjet, and $75 < \mJ < 105\GeV$ or $>$135\GeV.

Region A is the signal region, defined by the selection criteria described in the previous section. The multijet background yield in the signal region is obtained from regions B, C, and D, which are background enriched. Assuming that the \PQb-tagged subjet multiplicity and the Higgs boson mass are uncorrelated, the number of background events in the four regions follows the relationship:
\begin{equation}
	N_\text{A}/N_\text{C} = N_\text{B}/N_\text{D},
\end{equation}
where $N_\text{A}$, $N_\text{B}$, $N_\text{C}$, and $N_\text{D}$ are the yields in regions A, B, C, and D, respectively.
Thus, the number of background events in the signal region A is given by:
\begin{equation}
	N_\text{A} = N_\text{C} N_\text{B}/N_\text{D},
\end{equation}
after subtracting the \ttbar contribution predicted in the MC simulation. The contributions from \cPZ+jets and \PW+jets backgrounds are not subtracted as they are negligible.

The \mB~distribution of the multijet background in the signal region is estimated from the \MB~distribution in region C, since the reconstructed \MB~spectrum is not  expected to be correlated with the b jet multiplicity. The compatibility of the distributions in regions A and C is verified using simulated multijet events, and cross-checked in data.

In addition, the method is validated  using a signal-depleted region from sidebands at large mass. Here, two regions (A' and C') are defined, similar to A and C in the mass region $135 < \mJ < 165$\GeV. Two control regions (B' and D')  are defined requiring  $75 < \mJ < 105\GeV$ or $\mJ > 165\GeV$, respectively,  with 2 or 1 \PQb-tagged subjets.
The background distribution estimated in region A', using the method described above, agrees with the observed data in region A'. The systematic uncertainty in the normalization of the multijet background is taken to be equal to the observed difference between the predicted and the measured yields in region A'. It amounts to 10\% in the high purity category, and 5\% in the category with no forward jets.

\section{Systematic uncertainties} \label{syst}

The systematic effect of each source of uncertainty is evaluated by propagating the uncertainty in the input parameters to the reconstructed \PB~quark mass distribution and to the event yield. Then, the uncertainties in the event yield and in the \MB~distribution for signal and background processes are taken into account as ``nuisance'' parameters that are integrated over  in the statistical process of inferring the resultant parameters.

The statistical uncertainties in the background estimate of multijet production from control samples in data are propagated to \MB~in the signal region by changing the observed event yields in regions B and D, up and down by one standard deviation, and recalculating the expected distribution in the signal region.  As the expected multijet distribution in \MB~is estimated from region C, its statistical uncertainty in this region is considered in the signal extraction. In addition to the normalization, this uncertainty affects the distribution of the background \MB~in the  signal region. Therefore, a systematic uncertainty in the estimated shape of this distribution, arising from the limited number of events in the observed \MB~spectrum in region C, was derived by allowing the content of each bin to fluctuate independently according to Poisson statistics.

An additional systematic uncertainty in the estimated multijet background is obtained from the difference between the observed and predicted yields in the check, in the validation step that uses large-mass sideband regions, described in Section~\ref{bkg}, and corresponds to $\approx$5--10\%.

The systematic uncertainties from the limited number of simulated events and background estimates from simulation are also included by fluctuating each bin of the \MB~distribution independently, according to Poisson statistics.

Additional systematic uncertainties in simulated signal and background distributions originate from the corrections applied to rescale simulated distributions to data. Other such uncertainties are listed below. An uncertainty of 2.5\%~\cite{CMS:2017sdi} in the measured integrated luminosity is used just to account for the total event yields.

The corrections to account for the difference between the \PQb tagging efficiency measured in data and in simulation are changed up and down by their uncertainties in both AK4 jets and subjets. The reconstructed four-momenta of the AK4 and AK8 jets are also shifted by $\pm$1 standard deviation in the jet energy scale and resolution, and propagated to \MB. In addition, the pruned mass scale and resolution of the Higgs-tagged jet are changed within their uncertainties, affecting the \MB~spectrum by 0.5--5.5\%.

All simulated events are weighted to match the distribution of pileup interactions. The corresponding uncertainty is obtained by changing the total inelastic cross section by $\pm$4.6\%, which is used to calculate the pileup distribution in data. Scale factors are applied to account for differences between the trigger efficiency measured in data and in simulated events, with the uncertainties in the scale factors  applied as a function of \hT~and propagated to the \MB~distribution.

An additional uncertainty is applied to the simulated signal and backgrounds to account for discrepancies in the modelling of the forward jet multiplicity. The magnitude of this effect is obtained by considering the difference between the event yield in data and in MC, and results in an uncertainty of 0.5\% for the category with no forward jets, and 2.0\% for the category with at least one jet in the forward region.

The uncertainties from the choice of factorization and renormalization scales, $\mu_{\text F}$ and $\mu_{\text R}$, are taken into account by halving and doubling the nominal values and using the combination of $\mu_{\text F}$ and $\mu_{\text R}$ leading to the maximal change.
The resulting uncertainty in signal acceptance is as small as 1.3\%, depending on the mass hypothesis.  Larger effects (15--25\%) are observed in the overall normalization and acceptance in simulated backgrounds. In addition, the uncertainty from the choice of PDF is estimated by reweighting the simulated signal and background events using the NNPDF3.0~\cite{Botje:2011sn,Alekhin:2011sk,Ball2015} set of eigenvectors.

A summary of the systematic uncertainties considered in this analysis, along with their effect when propagated to the reconstructed \PB~mass, is presented in Table~\ref{tab:sistematics}.

\begin{table*}[hbt]
	\centering
	\topcaption{Summary of systematic uncertainties in background events. The quantification of the effects quoted in the table reflects the uncertainties in the event yields. All uncertainties are considered in the simulated background events, except the one on background estimation that affects only the data-based estimate of the multijet process. All the systematic uncertainties apply to both categories of forward-jet multiplicity, except for the case of the modelling of the forward jets, where the first entry corresponds to the category with no forward jets, and the second entry to the category with at least one jet in the forward region.}       \label{tab:sistematics}
      	\begin{tabular}{lc}
		\hline
        		Source & Effect \\
	        \hline
        		Luminosity & 2.5\%\\
	        \PQb tagging efficiency & 0--9\%\\
        		Misidentification efficiency & 0--2\% \\
	        Pileup modelling & 0--12\%\\
	        Trigger & $<$0.5\%\\
	        PDF & 1.0--4.5\% \\
        		$\mu_\text{R}$ and $\mu_\text{F}$ & 15--25\% \\
	        	Jet energy scale & 1--7\%\\
        		Jet energy resolution & 1.0--1.5\%\\
	        Jet mass scale & 0--5\%\\
        		Jet mass resolution & 0--4\%\\
	        MC Statistical accuracy & 1--4\%\\
        		Mismodelling of forward jets & 0.5/2.0\% \\
	        Background estimation & 5--10\%\\
		\hline
	\end{tabular}
\end{table*}

\section{Results} \label{results}

\begin{figure*}[hbtp]
	\centering
	\includegraphics[width=1.2\cmsFigWidth]{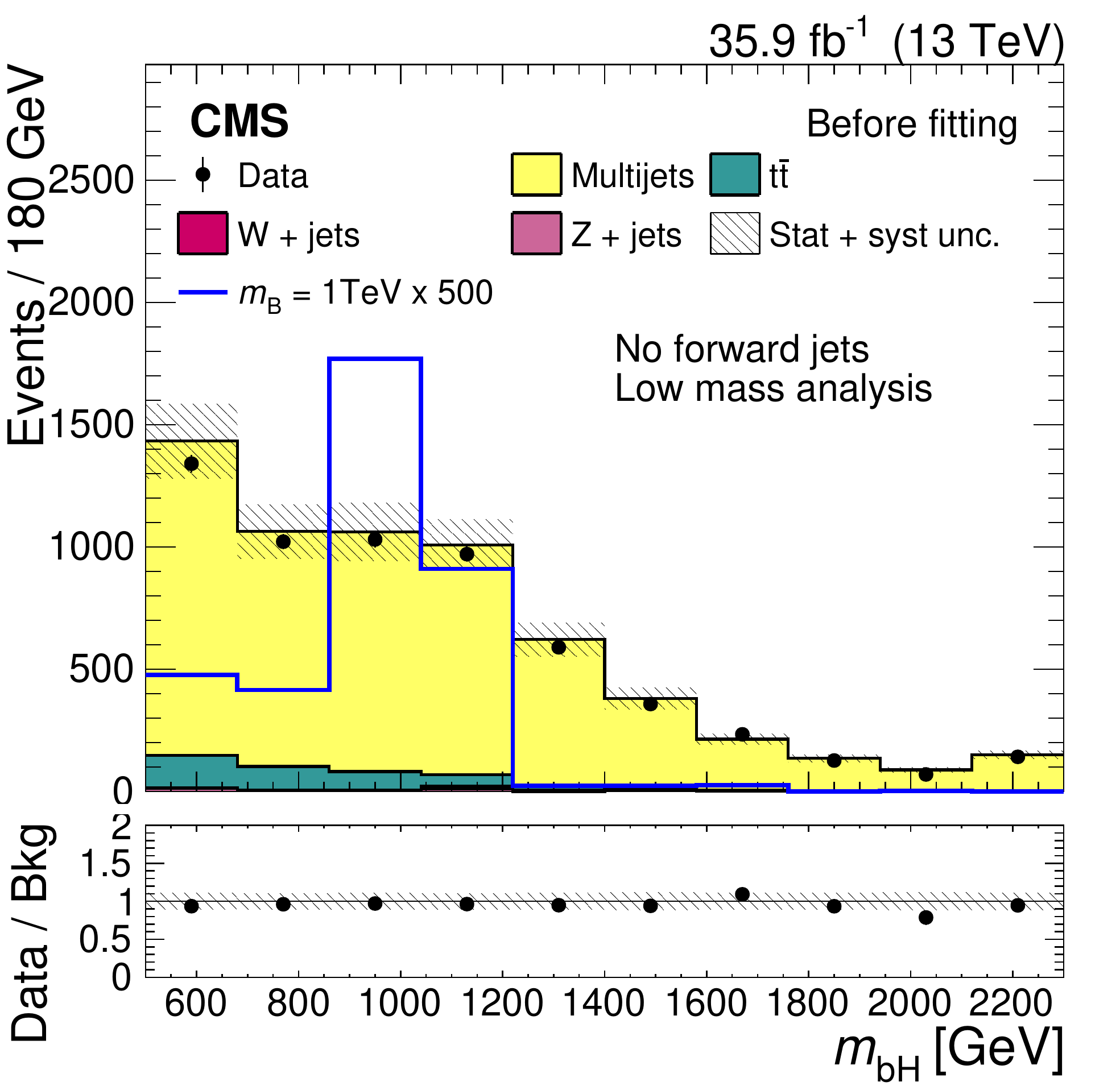}
	\includegraphics[width=1.2\cmsFigWidth]{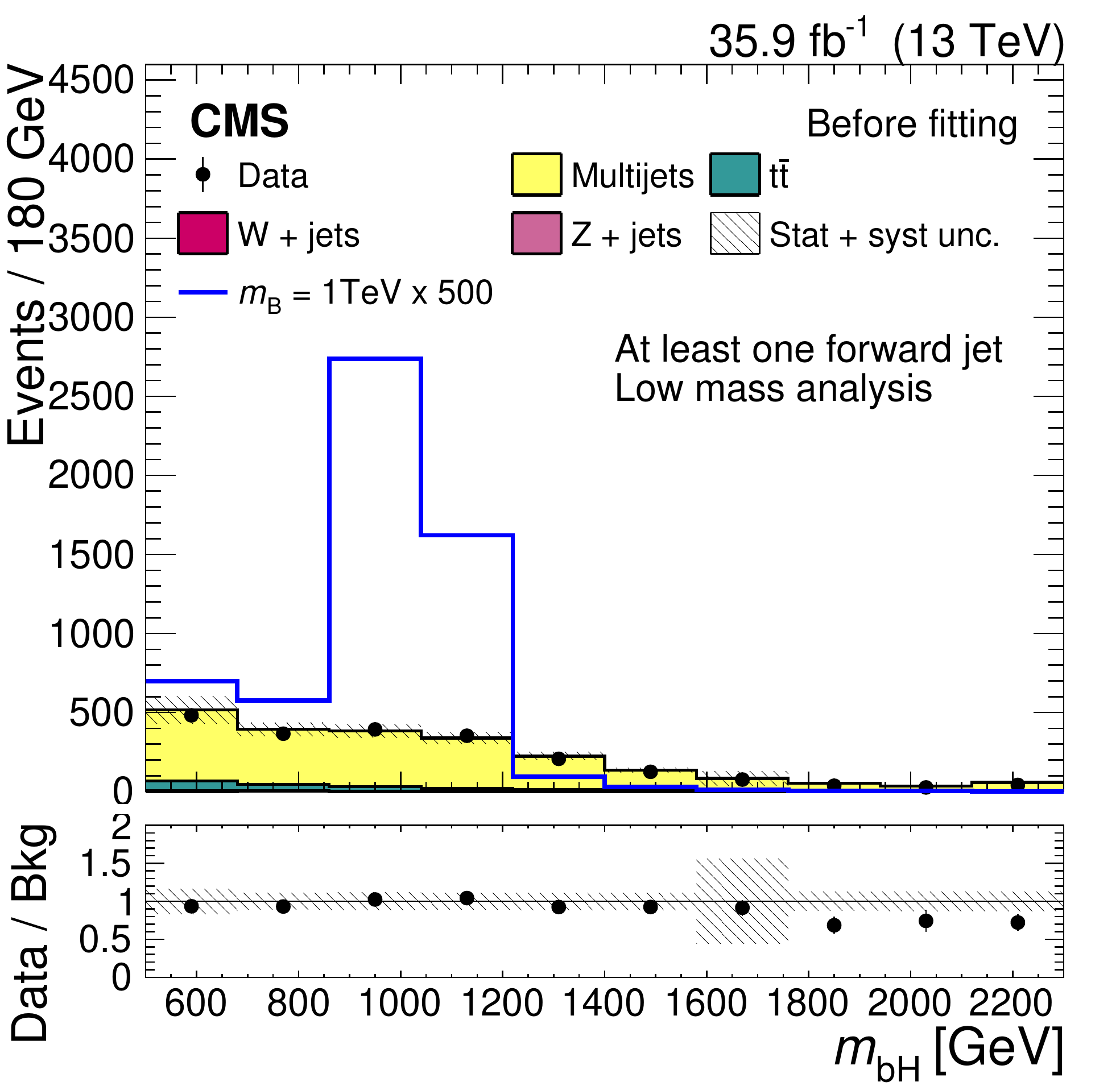}
	\includegraphics[width=1.2\cmsFigWidth]{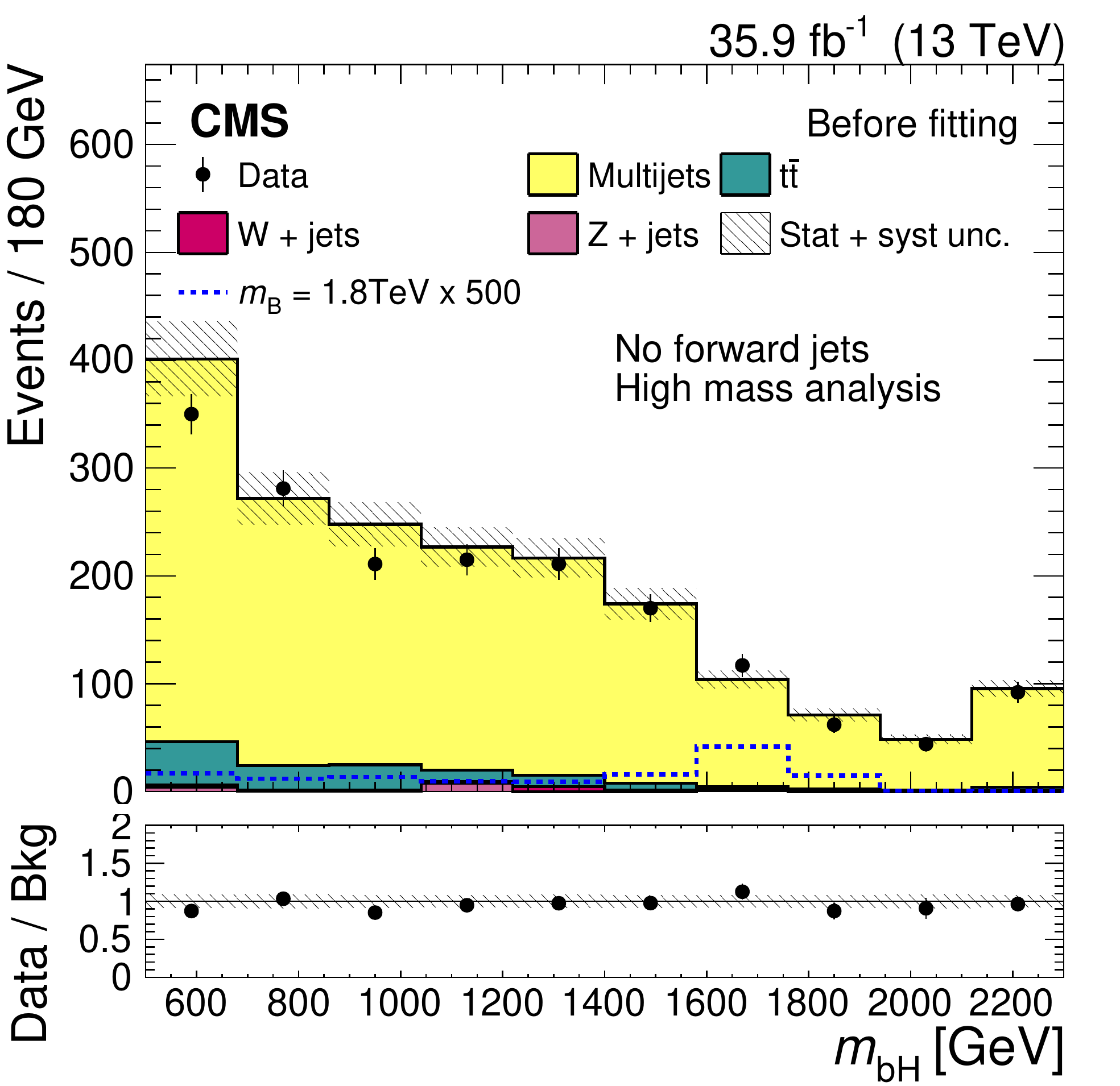}
	\includegraphics[width=1.2\cmsFigWidth]{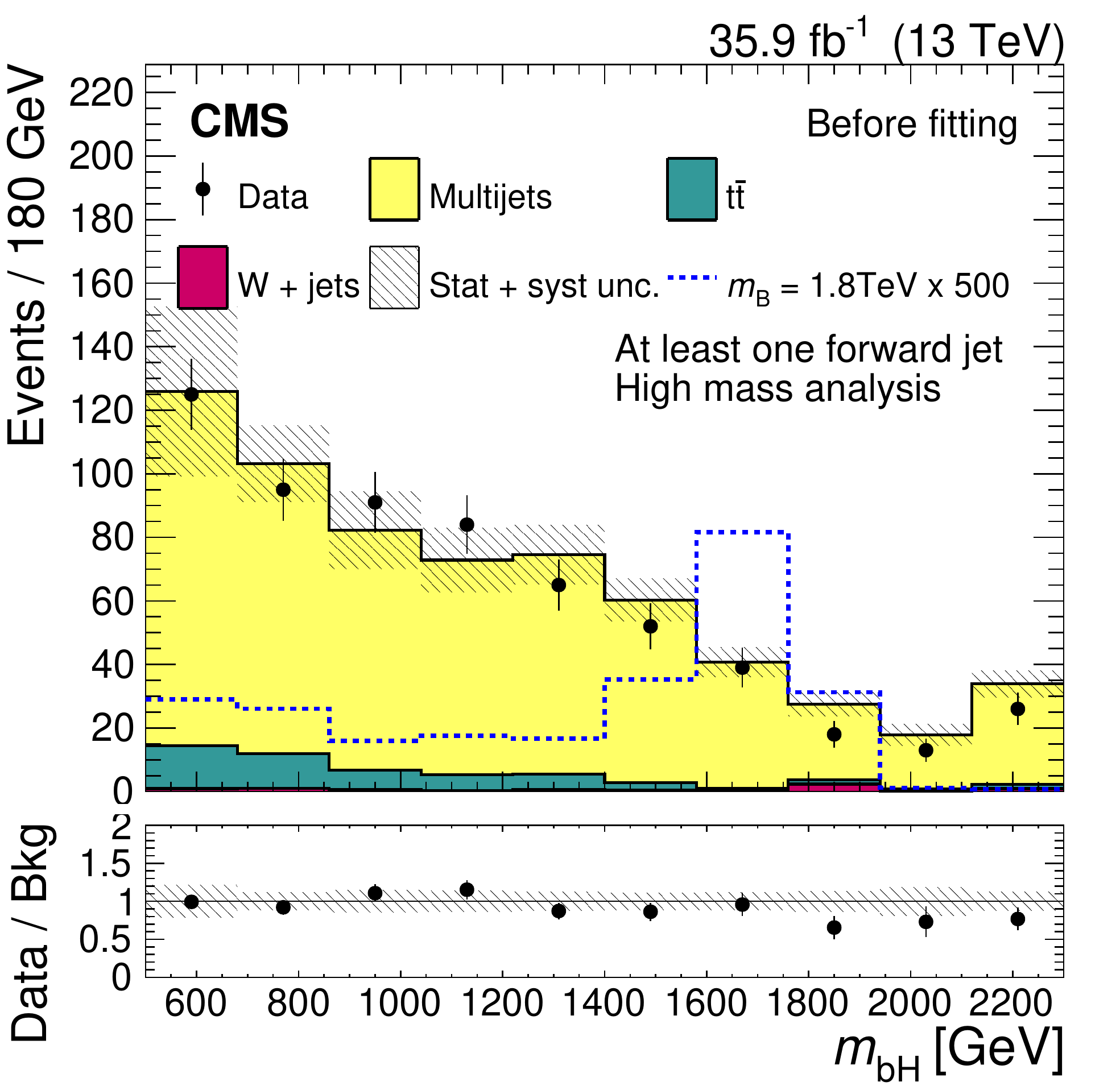}
	\caption{Distribution in the reconstructed B quark mass, after applying all selections to events with no forward jets (\cmsLeft) and to events with at least one forward jet (\cmsRight), compared to the background distributions estimated before fitting. The upper and lower plots refer to the low- and high-mass \mB~analyses, respectively. The expectations for signal MC events are given by the blue histogram lines. The different background contributions are indicated by the colour-filled histograms, and are obtained from Monte Carlo simulation, except for the multijets component, which is derived from data. The grey-hatched error band shows total uncertainties in the background expectation. The ratios of observations to background expectations are given in the lower panels, together with the total uncertainties prior to fitting, indicated by the grey-hatched band.}
	\label{fig:prebprimemass}
\end{figure*}

\begin{figure*}[hbtp]
	\centering
	\includegraphics[width=1.2\cmsFigWidth]{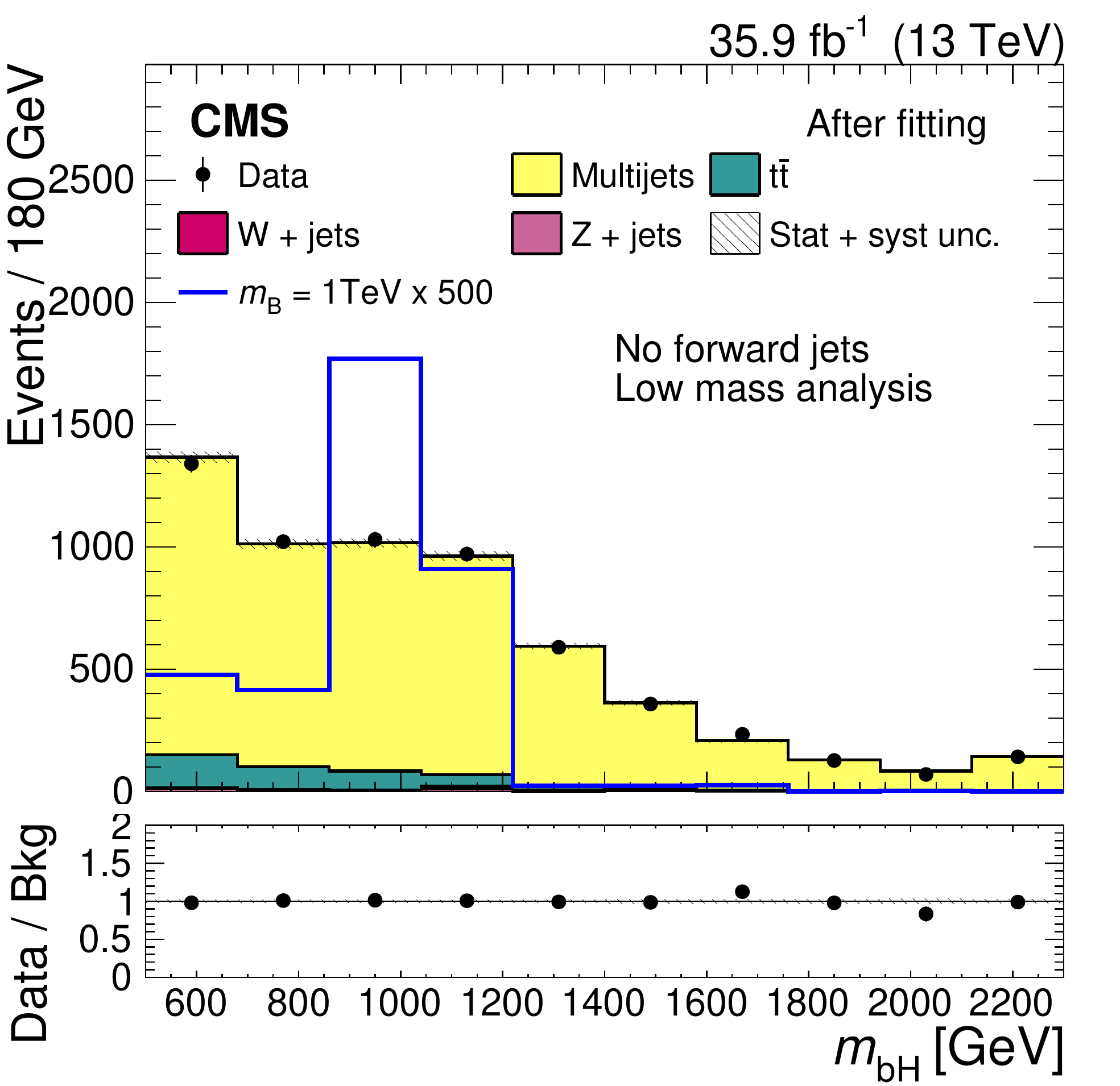}
	 \includegraphics[width=1.2\cmsFigWidth]{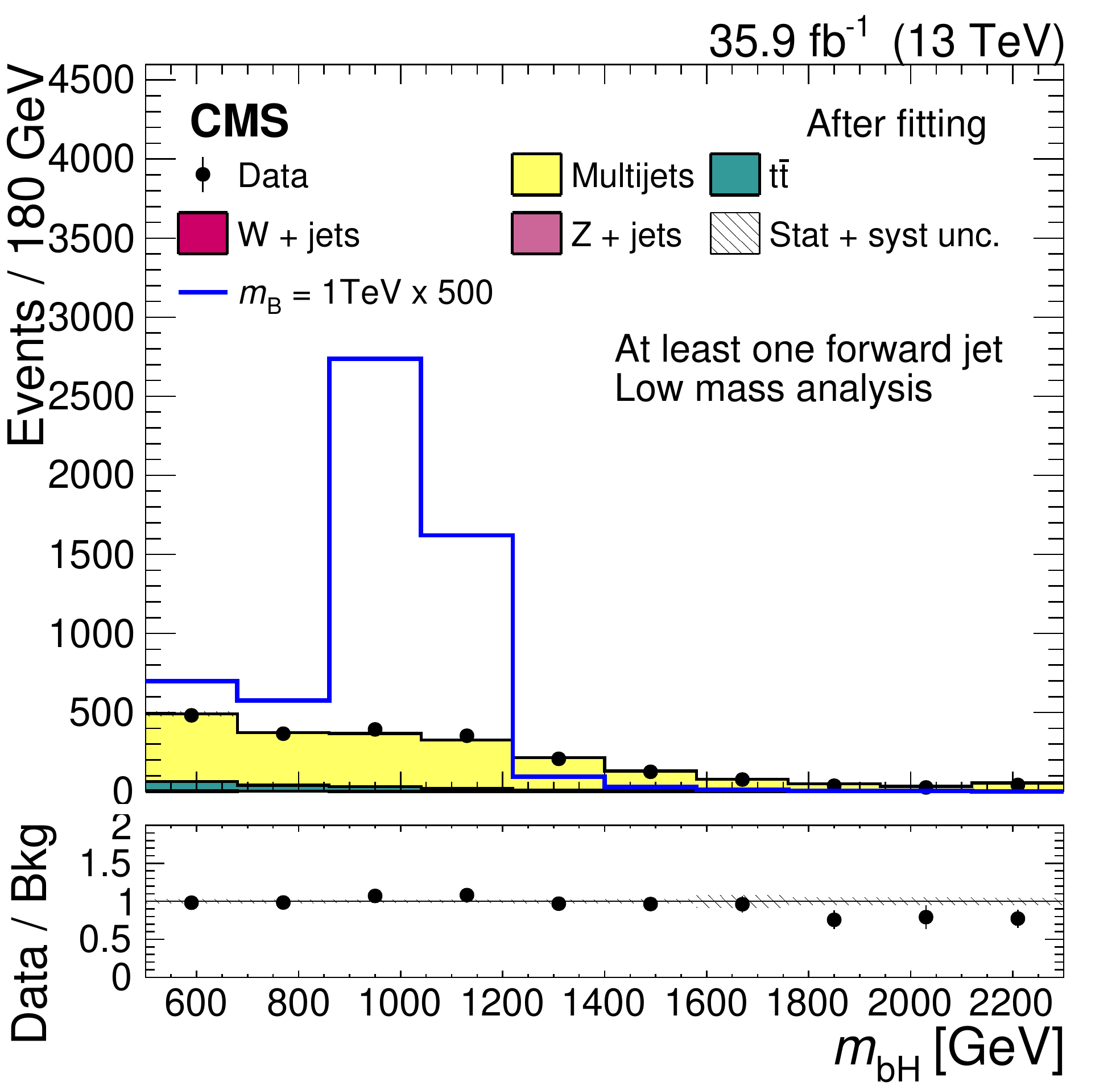}
	 \includegraphics[width=1.2\cmsFigWidth]{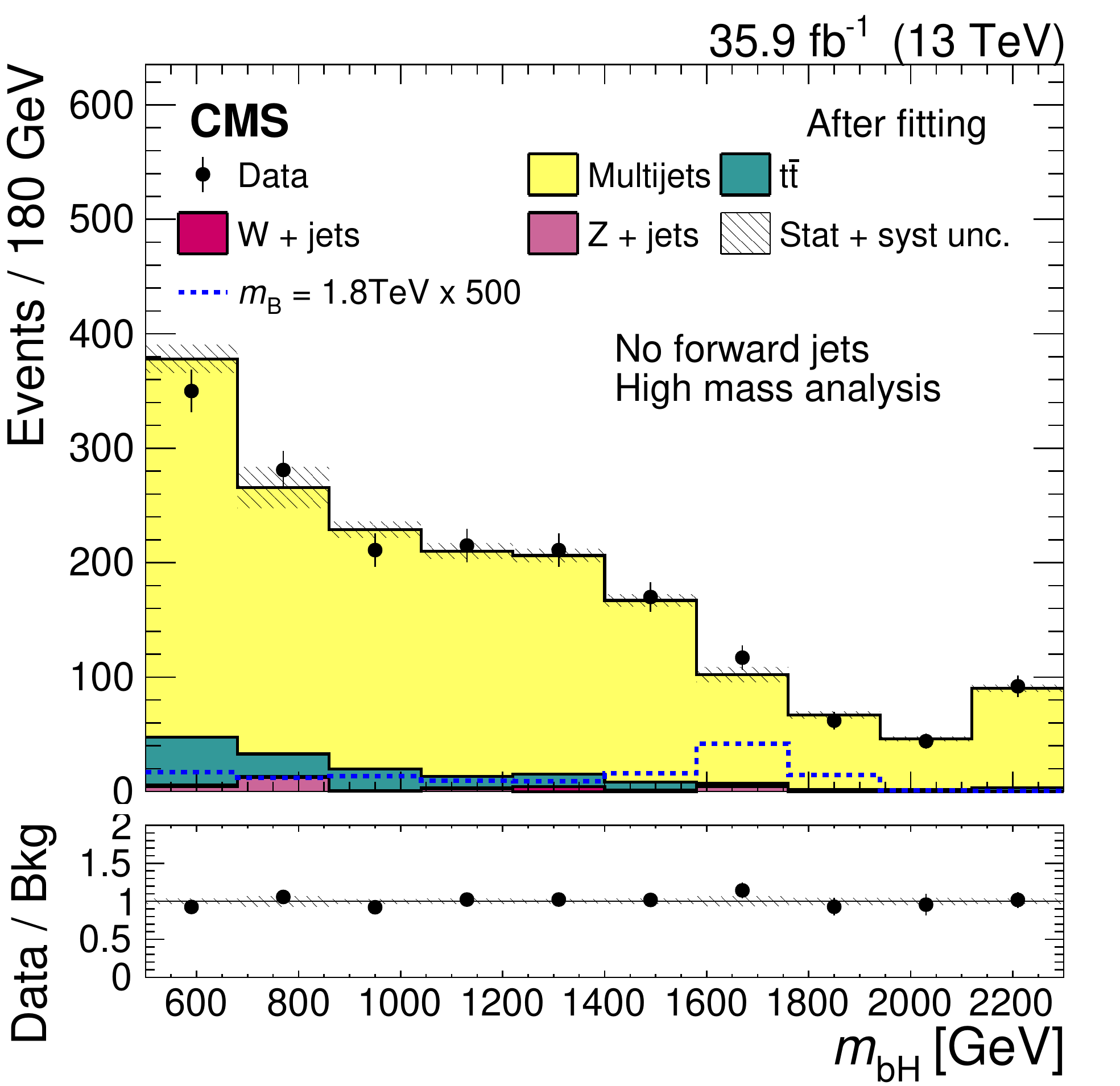}
	\includegraphics[width=1.2\cmsFigWidth]{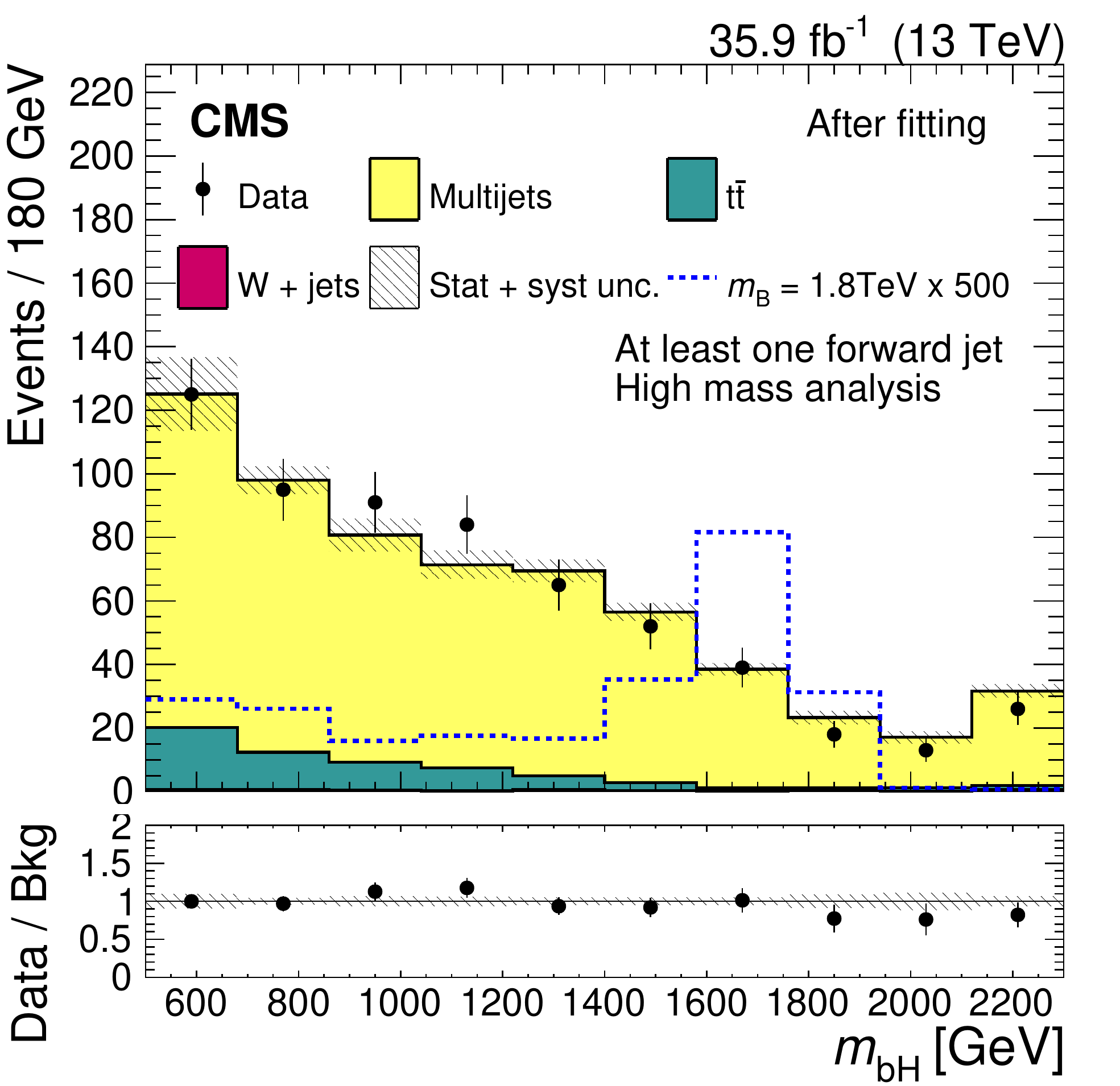}
    	\caption{Distribution in the reconstructed B quark mass after applying all selections to events with no forward jets (\cmsLeft) and to events with at least one forward jet (\cmsRight), compared to the background distributions estimated after fitting. The upper and lower plots refer to the low- and high-\mB~analyses, respectively. The expectations for signal MC events are given by the blue lines. The different background contributions are indicated by the colour-filled histograms, and are obtained from Monte Carlo simulation, except for the multijets component, which is derived from data. The grey-hatched error band shows total uncertainties in the background expectation. The ratios of the observations to background expectations are given in the lower panels, together with the total uncertainties after fitting, indicated by the grey-hatched band.}
    	\label{fig:postbprimemass}
\end{figure*}

A binned maximum likelihood fit is performed to the \MB~distribution in Fig.~\ref{fig:prebprimemass}, where the dominant multijet background is estimated from data, as discussed in Section~\ref{bkg}. The fitted \MB~distributions are presented in Fig.~\ref{fig:postbprimemass}, while the expected yields are listed in Table~\ref{tab:sigyields} for the backgrounds, and for two signal hypotheses ($\MB = 1000$ and 1800\GeV), together with their observed yields. The observed distributions are consistent with the background-only hypothesis in all the categories. Upper limits are set therefore on the product of the cross section and branching fraction of a \PB~quark decaying to \PH{}\PQb, produced in association with another \PQb~quark and a light-flavoured quark, as a function of \MB. Exclusion limits at 95\% confidence level (\CL) are calculated using a modified frequentist approach and a profile likelihood ratio as test statistic, in an asymptotic approximation~\cite{Junk:1999kv,Read:2002hq,Cowan:2010js}. The combination of the two forward-jet multiplicity-based categories increases the sensitivity of the analysis by up to 20\% relative to that obtained when only requiring at least one jet in the $\abs{\eta}>2.4$ region of the detector.

Systematic uncertainties described in Section~\ref{syst} are treated as nuisance parameters affecting the rate of the expected \MB~distribution. Both the uncertainties affecting the normalization, modelled using log-normal priors, and uncertainties in distributions are included in the fit~\cite{Conway:2011in}.

\begin{figure*}[hbtp]
	\includegraphics[width=1.2\cmsFigWidth]{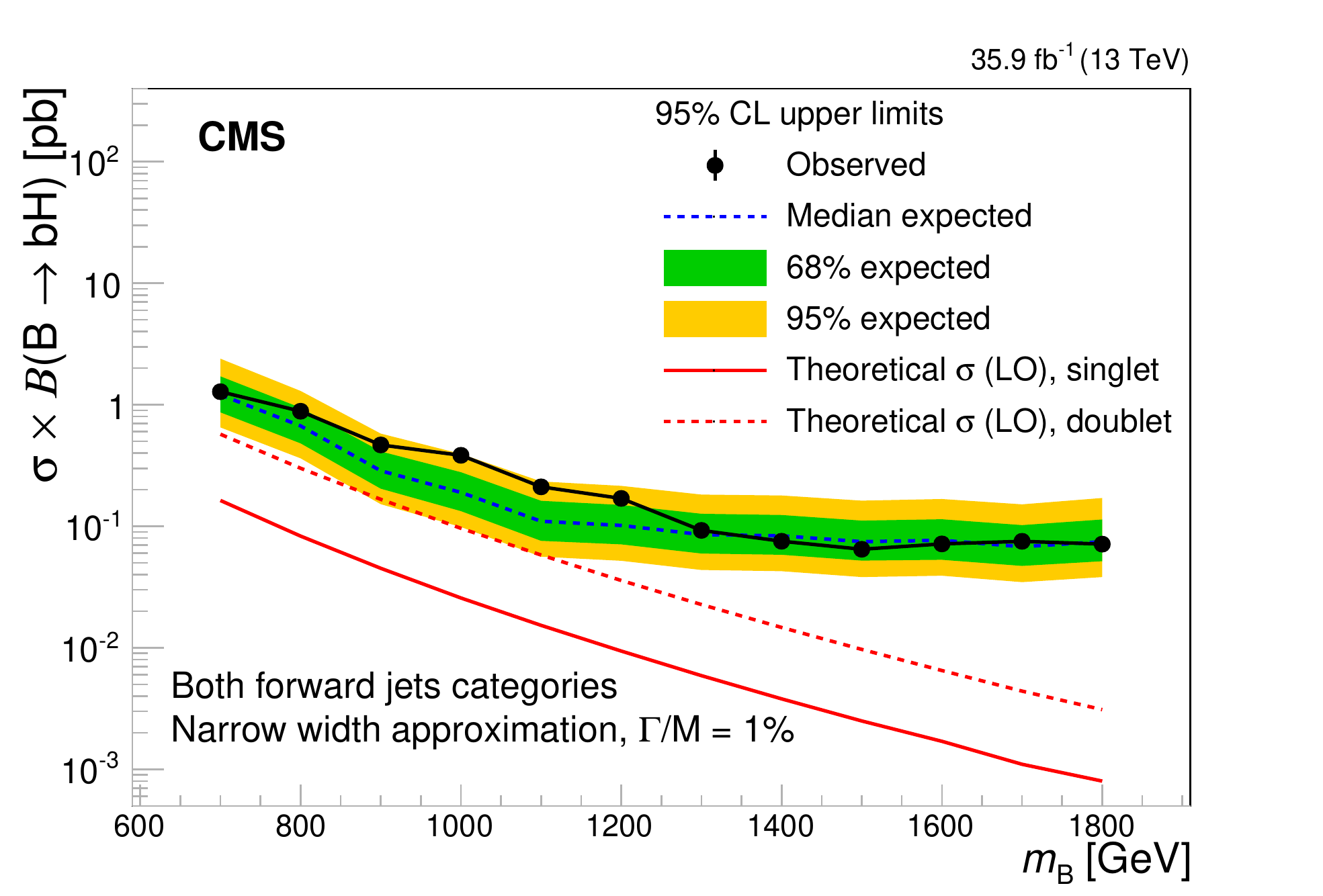}
    	\includegraphics[width=1.2\cmsFigWidth]{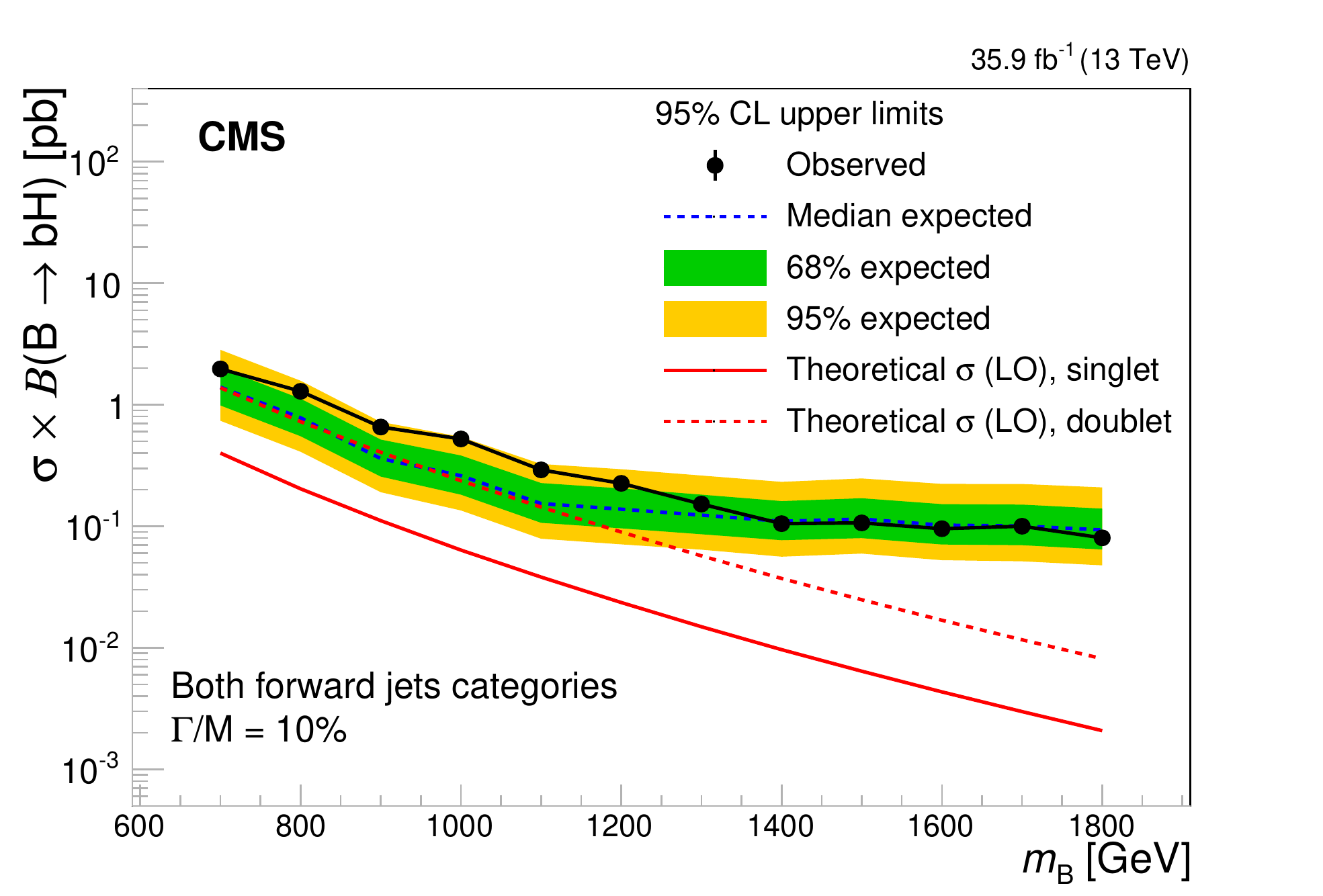}
    	\includegraphics[width=1.2\cmsFigWidth]{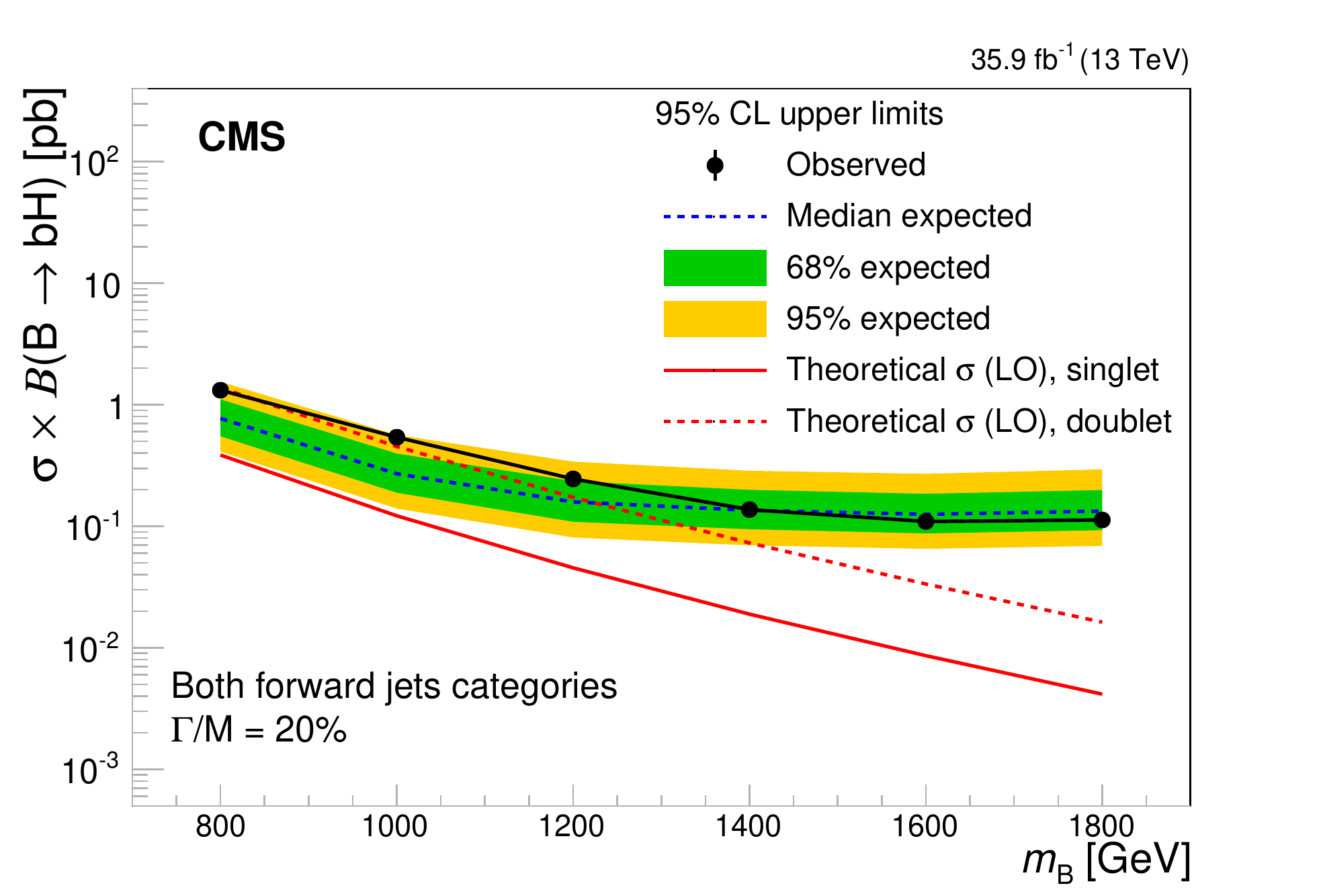}
    	\includegraphics[width=1.2\cmsFigWidth]{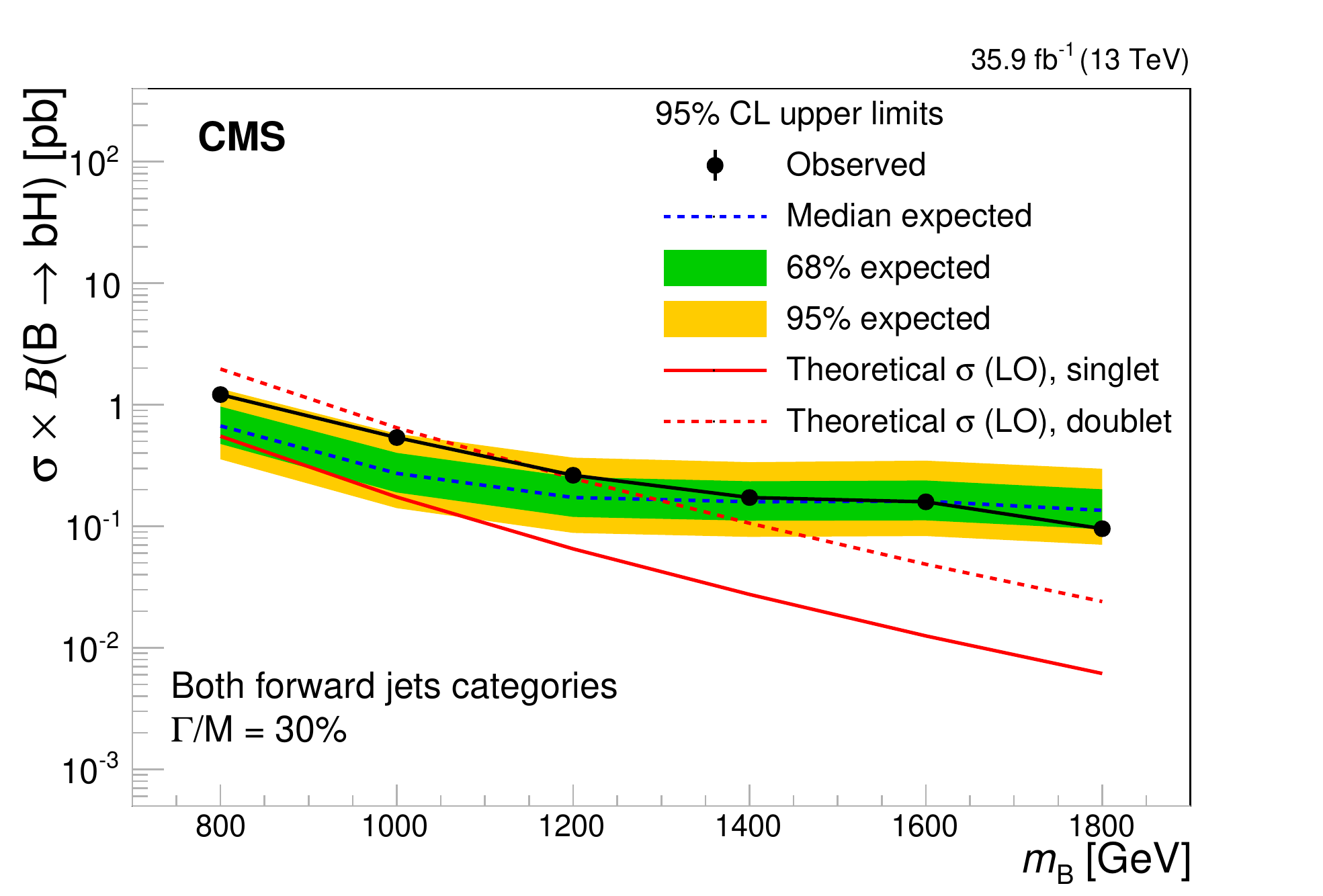}
    	\caption{The median observed and expected 95\% \CL upper limits on the product of the B quark production cross section  and branching fraction as a function of the signal mass, assuming narrow-width resonances (upper-left) and widths of 10 (upper-right), 20 (lower-left), and 30\% (lower-right) of the resonance mass for the B quark. The results are shown for the combination of 0 and $>$0 forward-jet categories. The continuous red curves correspond to the theoretical expectations for singlet and doublet models.}
    	\label{fig:limits_ww}
\end{figure*}

The observed and expected combined upper limits from the two categories are given in Fig.~\ref{fig:limits_ww}. Assuming a narrow width, values of $\sigma \, \mathcal{B}{(\PH\PQb)}$ between 0.07--1.28\unit{pb} are excluded at the 95\% confidence level, for masses in the range 700--1800\GeV. Upper limits are compared with the predictions calculated at NLO~\cite{Matsedonskyi:2014mna} for both singlet and doublet B quark models, assuming narrow widths and $\mathcal{B}{(\PH\PQb)} \approx 25$\%.
Figure~\ref{fig:limits_ww} also shows the observed and expected upper limits on the product of the cross section and branching fraction for \PB~quarks with intrinsic widths fixed to $\Gamma_{\PB}/\mB = 10$, 20, and 30\%. Sensitivities similar to those for negligible widths are observed for exclusion limits that lie between 0.08 and 1.97, 0.11 and 1.32, and 0.10 and 1.22\unit{pb}, respectively, for the 10, 20, and 30\% $\Gamma_{\PB}/$\mB~values.

\begin{table*}[hbt]
	\centering
	\topcaption{Observed and expected fitted number of events in the signal ranges of $700 < \mB < 1500$ and $1500 < \mB < 1800\GeV$, and expected signal at $\mB = 1000$ and 1800\GeV. The multijet background is obtained from data, while the yields for the other sources of background are obtained from MC simulation. The combined statistical and systematic uncertainties correspond to the quadrature of the statistical and systematic uncertainties.}
      	\label{tab:sigyields}
      	\begin{tabular}{c|ccc}
      		\hline
		{Category} & {Source} & {$700 < \mB < 1500\GeV$} & {$1500 < \mB < 1800\GeV$}\\\hline
		\multirow{7}{*}{No forward jets}& \ttbar & $394 \pm 46$ & $117 \pm 18$\\
		& \PW+jets & $29 \pm 13$ & $10.5 \pm 4.3$\\
		& \cPZ+jets & $43 \pm 15$ & $23 \pm 23$\\
		& Multijets  & $5416 \pm 60$ & $1612 \pm 24$\\[\cmsTabSkip]
		& Total background & $5882 \pm 42$ & $1762 \pm 26$\\[\cmsTabSkip]
		& Observed in data & $5886 \pm 77$ & $1753 \pm 42$\\[\cmsTabSkip]
		& Expected signal & $7.3 \pm 0.3$ & $0.27 \pm 0.01$\\
		\hline
		\multirow{7}{*}{$>$0 forward jets}& \ttbar & $163 \pm 20$ & $58 \pm 17$\\
		& \PW+jets & $11.5 \pm 4.2$ & $4.3 \pm 1.4$\\
		& \cPZ+jets & $2^{+10}_{-2}$ & $-$\\
		& Multijets & $1938 \pm 23$ & $549 \pm 10$\\[\cmsTabSkip]
		& Total background & $2115 \pm 21$ & $612 \pm 15$\\[\cmsTabSkip]
		& Observed in data & $2107 \pm 46$ & $608 \pm 25$\\[\cmsTabSkip]
		& Expected signal & $11.5 \pm 0.3$ & $0.51 \pm  0.01$\\
		\hline
	\end{tabular}
\end{table*}

\clearpage

\section{Summary} \label{summary}

A search has been presented for electroweak production of vector-like \PB~quarks with charge $-1/3\,e$, decaying to a bottom quark and a Higgs boson (\PH). The analysis uses a data sample corresponding to an integrated luminosity of \intLumi, collected in \Pp\Pp~collisions at $\sqrt{s} = 13\TeV$.

No significant deviations are observed relative to the standard model prediction, and upper limits are placed on the product of the  cross section and the branching fraction of the \PB~quark.

Expected and observed limits at 95\% confidence level vary from 1.20 to 0.07\unit{pb} and from 1.28 to 0.07\unit{pb}, respectively, for \PB~quark masses in the range considered, which extends from 700 to 1800\GeV.
The search is performed under the hypothesis of a singlet or doublet B quark of narrow width decaying to \PH{}\PQb~with a branching fraction of approximately 25\%. The possibility of having  non-negligible resonant widths is also studied. Limits obtained on the production of \PB~quarks with  widths of 10, 20, and 30\% of the resonance mass are comparable to those found for the narrow-width approximation. This search extends existing knowledge on vector-like quarks, by interpreting the results in a new theoretical framework with non-negligible resonance widths, and investigating the final state with a bottom quark and a Higgs boson for the first time.

\begin{acknowledgments}
We congratulate our colleagues in the CERN accelerator departments for the excellent performance of the LHC and thank the technical and administrative staffs at CERN and at other CMS institutes for their contributions to the success of the CMS effort. In addition, we gratefully acknowledge the computing centers and personnel of the Worldwide LHC Computing Grid for delivering so effectively the computing infrastructure essential to our analyses. Finally, we acknowledge the enduring support for the construction and operation of the LHC and the CMS detector provided by the following funding agencies: BMWFW and FWF (Austria); FNRS and FWO (Belgium); CNPq, CAPES, FAPERJ, and FAPESP (Brazil); MES (Bulgaria); CERN; CAS, MoST, and NSFC (China); COLCIENCIAS (Colombia); MSES and CSF (Croatia); RPF (Cyprus); SENESCYT (Ecuador); MoER, ERC IUT, and ERDF (Estonia); Academy of Finland, MEC, and HIP (Finland); CEA and CNRS/IN2P3 (France); BMBF, DFG, and HGF (Germany); GSRT (Greece); OTKA and NIH (Hungary); DAE and DST (India); IPM (Iran); SFI (Ireland); INFN (Italy); MSIP and NRF (Republic of Korea); LAS (Lithuania); MOE and UM (Malaysia); BUAP, CINVESTAV, CONACYT, LNS, SEP, and UASLP-FAI (Mexico); MBIE (New Zealand); PAEC (Pakistan); MSHE and NSC (Poland); FCT (Portugal); JINR (Dubna); MON, RosAtom, RAS, RFBR and RAEP (Russia); MESTD (Serbia); SEIDI, CPAN, PCTI and FEDER (Spain); Swiss Funding Agencies (Switzerland); MST (Taipei); ThEPCenter, IPST, STAR, and NSTDA (Thailand); TUBITAK and TAEK (Turkey); NASU and SFFR (Ukraine); STFC (United Kingdom); DOE and NSF (USA).

\hyphenation{Rachada-pisek} Individuals have received support from the Marie-Curie program and the European Research Council and Horizon 2020 Grant, contract No. 675440 (European Union); the Leventis Foundation; the A. P. Sloan Foundation; the Alexander von Humboldt Foundation; the Belgian Federal Science Policy Office; the Fonds pour la Formation \`a la Recherche dans l'Industrie et dans l'Agriculture (FRIA-Belgium); the Agentschap voor Innovatie door Wetenschap en Technologie (IWT-Belgium); the F.R.S.-FNRS and FWO (Belgium) under the "Excellence of Science - EOS" - be.h project n. 30820817; the Ministry of Education, Youth and Sports (MEYS) of the Czech Republic; the Council of Science and Industrial Research, India; the HOMING PLUS program of the Foundation for Polish Science, cofinanced from European Union, Regional Development Fund, the Mobility Plus program of the Ministry of Science and Higher Education, the National Science Center (Poland), contracts Harmonia 2014/14/M/ST2/00428, Opus 2014/13/B/ST2/02543, 2014/15/B/ST2/03998, and 2015/19/B/ST2/02861, Sonata-bis 2012/07/E/ST2/01406; the National Priorities Research Program by Qatar National Research Fund; the Programa Severo Ochoa del Principado de Asturias; the Thalis and Aristeia programs cofinanced by EU-ESF and the Greek NSRF; the Rachadapisek Sompot Fund for Postdoctoral Fellowship, Chulalongkorn University and the Chulalongkorn Academic into Its 2nd Century Project Advancement Project (Thailand); the Welch Foundation, contract C-1845; and the Weston Havens Foundation (USA).
\end{acknowledgments}

\bibliography{auto_generated}
\cleardoublepage \appendix\section{The CMS Collaboration \label{app:collab}}\begin{sloppypar}\hyphenpenalty=5000\widowpenalty=500\clubpenalty=5000\textbf{Yerevan Physics Institute,  Yerevan,  Armenia}\\*[0pt]
A.M.~Sirunyan, A.~Tumasyan
\vskip\cmsinstskip
\textbf{Institut f\"{u}r Hochenergiephysik,  Wien,  Austria}\\*[0pt]
W.~Adam, F.~Ambrogi, E.~Asilar, T.~Bergauer, J.~Brandstetter, E.~Brondolin, M.~Dragicevic, J.~Er\"{o}, A.~Escalante Del Valle, M.~Flechl, M.~Friedl, R.~Fr\"{u}hwirth\cmsAuthorMark{1}, V.M.~Ghete, J.~Grossmann, J.~Hrubec, M.~Jeitler\cmsAuthorMark{1}, A.~K\"{o}nig, N.~Krammer, I.~Kr\"{a}tschmer, D.~Liko, T.~Madlener, I.~Mikulec, E.~Pree, N.~Rad, H.~Rohringer, J.~Schieck\cmsAuthorMark{1}, R.~Sch\"{o}fbeck, M.~Spanring, D.~Spitzbart, A.~Taurok, W.~Waltenberger, J.~Wittmann, C.-E.~Wulz\cmsAuthorMark{1}, M.~Zarucki
% ...
\vskip\cmsinstskip
% ...
1:~~Also at Vienna University of Technology, Vienna, Austria\\
% ...

\end{sloppypar}
\end{document}